\documentclass{jfm}
\usepackage{graphicx}
\usepackage{epstopdf, epsfig}
\usepackage[utf8]{inputenc}
\usepackage{comment}

\usepackage{color}
\usepackage{amsmath}
\usepackage{amssymb}
\usepackage[version=3]{mhchem}
\usepackage{mathrsfs}      
\usepackage{booktabs}   
\usepackage{subfigure}
\usepackage{float}
\usepackage{appendix}
\usepackage{comment}

\usepackage{epsfig}
\usepackage{dcolumn}%
\usepackage{soul}

\title{A General Drag Coefficient for Flow over a Sphere}

\author{Narendra Singh \aff{1}
  \corresp{\email{narsingh@stanford.edu}},
  Michael Kroells \aff{2},
  Chenxi Li \aff{3},
Eric Ching \aff{1},
Matthias Ihme  \aff{1},
Christopher J. Hogan \aff{3},
 \and Thomas Schwartzentruber \aff{2}}

\affiliation{\aff{1}Department of Mechanical Engineering, Stanford University, Stanford, CA 94305, USA
\aff{2}Department of Aerospace Engineering and Mechanics, University of Minnesota, Minneapolis, Minnesota 55455, USA,
\aff{3} Department of Mechanical Engineering, University of Minnesota, Minneapolis, Minnesota 55455, USA
}

\begin{document}

\maketitle

\begin{abstract}
A generalized physics-based expression for the drag coefficient of spherical particles moving in a fluid is derived. The proposed correlation incorporates essential rarefied physics, low-speed hydrodynamics, and shock-wave physics to
accurately model the particle-drag force for a wide range of Mach and Knudsen numbers (and therefore Reynolds number) a particle may experience. Owing to the basis of the derivation in physics-based scaling laws, the proposed correlation embeds gas-specific properties and has explicit dependence on the ratio of specific heat capacities at constant pressure and constant volume. The correlation is applicable for arbitrary particle relative velocity, particle diameter, gas pressure, gas temperature, and surface temperature.  Compared to existing drag models, the correlation is shown to more accurately reproduce a wide range of experimental data. Finally, the new correlation is applied to simulate dust particles' trajectories in high-speed flow, relevant to a spacecraft entering the Martian atmosphere. The enhanced surface heat flux due to particle impact is found to be sensitive to the particle drag model. 
\end{abstract}

\begin{keywords}
Authors should not enter keywords on the manuscript, as these must be chosen by the author during the online submission process and will then be added during the typesetting process (see http://journals.cambridge.org/data/\linebreak[3]relatedlink/jfm-\linebreak[3]keywords.pdf for the full list)
\end{keywords}

\section{Introduction}
An understanding of particle and droplet migration in high-speed aerospace flows is important in quantitatively describing a number of engineered systems.  In particular, for high-speed  flight vehicles, dust particles, water droplets in the atmosphere, or spalling fibers from the vehicle's surface may alter surface heating rates \citep{ching2020sensitivity} and erode the heat shield's surface. This is particularly relevant for future Mars and Earth entry missions, where particulate matter due to dust storms\citep{Mars_dust}, droplets or ice crystals may be present.  Particle based coating processes, including cold spray deposition \citep{SCHMIDT2006},  plasma spray deposition \citep{herman1988}, and aerosol deposition \citep{Akedo2006}, all involve the acceleration of particles to supersonic speeds, followed by inertial impaction of the desired particles with a substrate.  The quality and resulting properties of the coating are strongly dependent on the particle size and particle impaction velocity.  In monitoring particle trajectories in high speed environments, accurate drag force calculations are essential; finite particle inertia yields appreciable velocity differences between particles and the surroundings at shock fronts and near surfaces.  However, even if high speed gas flows are in the continuum regime, the flow relative to the motion of the particle may be in the rarefied regime.  Building upon prior theoretical work in understanding drag on a spherical particle in low to high speed, and continuum to rarefied flows, the purpose of this work is to develop a physics-based expression for the drag coefficient, applicable under very general conditions, i.e. over a wide range of Mach ($M_{\infty}$) number and Knudsen ($Kn_\infty$) number (and hence Reynolds number $Re_\infty$ $\propto$ $M_{\infty}$/$Kn_\infty$).       

 In principle, drag force ($F_d$) on a particle of radius $R$ depends on the free-stream relative velocity between the particle and the fluid ($U_{\infty}$), density  ($\rho_{\infty}$), viscosity ($\mu_{\infty}$), temperature ($T_{\infty}$) of the fluid, and surface temperature ($T_w$) of the particle. The coefficient of drag is defined as:
 \begin{equation}
    C_d = \cfrac{F_d}{\cfrac{1}{2} \rho_\infty U_{\infty}^2 \pi R^2}
\end{equation}
 While for restricted ranges of $Re_{\infty}$, $M_{\infty}$ or $Kn_{\infty}$, theoretical estimation of drag correlation is possible, generalizations have relied on \textit{ad-hoc} interpolations \citep{Henderson,loth2008compressibility} between different regimes or neural network based empirical formulations \citep{li2019mass}.  Intermediate regimes encompass a range of physical mechanisms, such as inertial effects at high $Re_\infty$, viscous effects at low $Re_\infty$, strong gas compression across shock-waves in high-speed (or high $M_\infty$) flows, and non-continuum effects due to fewer inter-molecular collisions and wall-gas collisions in the rarefied regime ($Kn_\infty>1.0$). In this work, we develop a physics-based generalized drag correlation by successively incorporating the contribution of these complex mechanisms through consistent physical analysis. The parameters introduced in the scaling laws are then obtained using available experimental measurements and first-principles based simulation data.  One feature of the proposed correlation is its explicit dependence on the gas-type, specifically on the ratio of specific heat capacities at constant pressure and volume, which is absent in state-of-the-art correlations. We apply the proposed correlation to simulate high-speed dusty flow over a sphere to show that the surface heating rate is sensitive to the drag correlation used for estimating trajectories of dust particles. The current formulation is only for spherical objects, however, generalizations to other shapes are possible \citep{zhang2012determination}.
 
\section{\label{sec:level2}Theory and Approach}
Physical mechanisms and their relative contribution to the overall drag may vary depending on the flow regime. For instance, viscous forces dominate drag force for low $Re_\infty$, while inertial forces dominate as $Re_\infty$ increases. In the present work, the flow regimes are divided based on compressibility and rarefaction effects. The Mach number ($M_\infty$) demarcates regimes based on compressibility effects such as incompressible regime ($M_\infty<0.3$), compressible ($0.3<M_\infty<1.0$), supersonic ($M_\infty>1.0$), and hypersonic regime ($M_\infty>5.0$). Similarly, rarefaction effects are characterized by the Knudsen number ($Kn_\infty$), which is defined as the ratio of mean free path to the characteristic length scale ($2R$). $Kn_\infty$ can also be expressed in terms of $M_\infty$ and $Re_\infty$ as:
\begin{equation}
    {Kn_\infty} = \cfrac{M_\infty}{Re_\infty} \sqrt{\cfrac{\gamma \pi}{2}}
    \label{kn_def}
\end{equation}
Low $Kn$ ($Kn<0.01$) corresponds to the continuum regime, followed by slip regime ($0.01<Kn<0.1$), transition regime ($1<Kn<10$) and free-molecular regime ($Kn>10$). In the following subsections, the aforementioned flow regimes are considered and physics-based scaling laws for a generalized drag correlation are derived. 
\subsection{Continuum Regime Formulation}
In this section, we consider the continuum regime ($Kn<0.01$) for any $M_{\infty}$, provided that $Re_{\infty}$ remains less than the critical transition condition, 
$Re_{\infty,c}\leq 10^4$  (i.e. before the onset of boundary layer turbulence effects). In terms of organization, incompressible flow is considered  first, followed by compressible flow, which is further divided into subsonic, supersonic, and hypersonic regimes.

   \begin{figure}
   \centering
  \subfigure[Incompressible flow]
  {
    \includegraphics[width=0.4\linewidth]{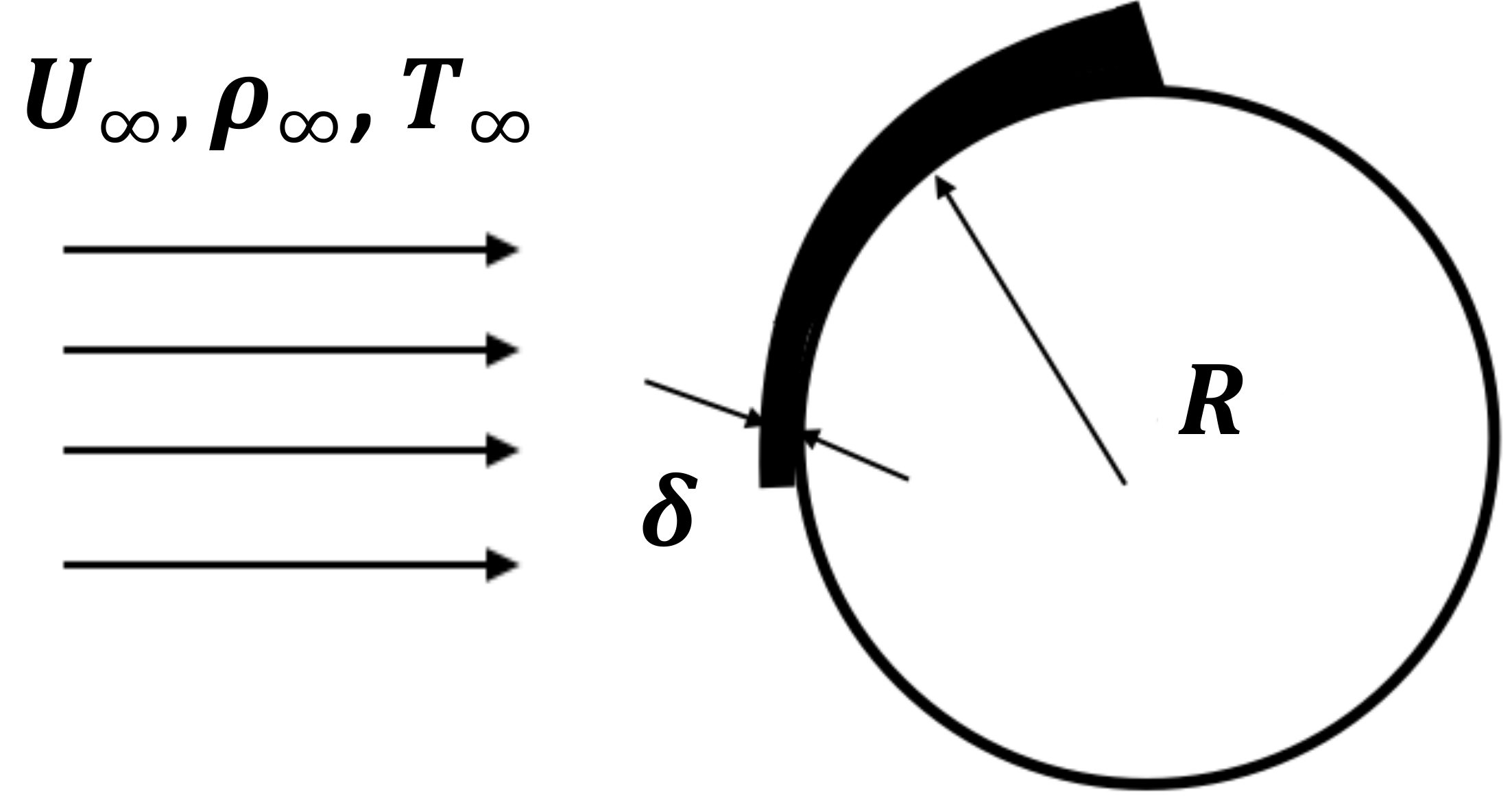}
   \label{incompressible_sphere}
   }  
  \subfigure[Compressible flow (subsonic)]
  {
    \includegraphics[width=0.50\linewidth]{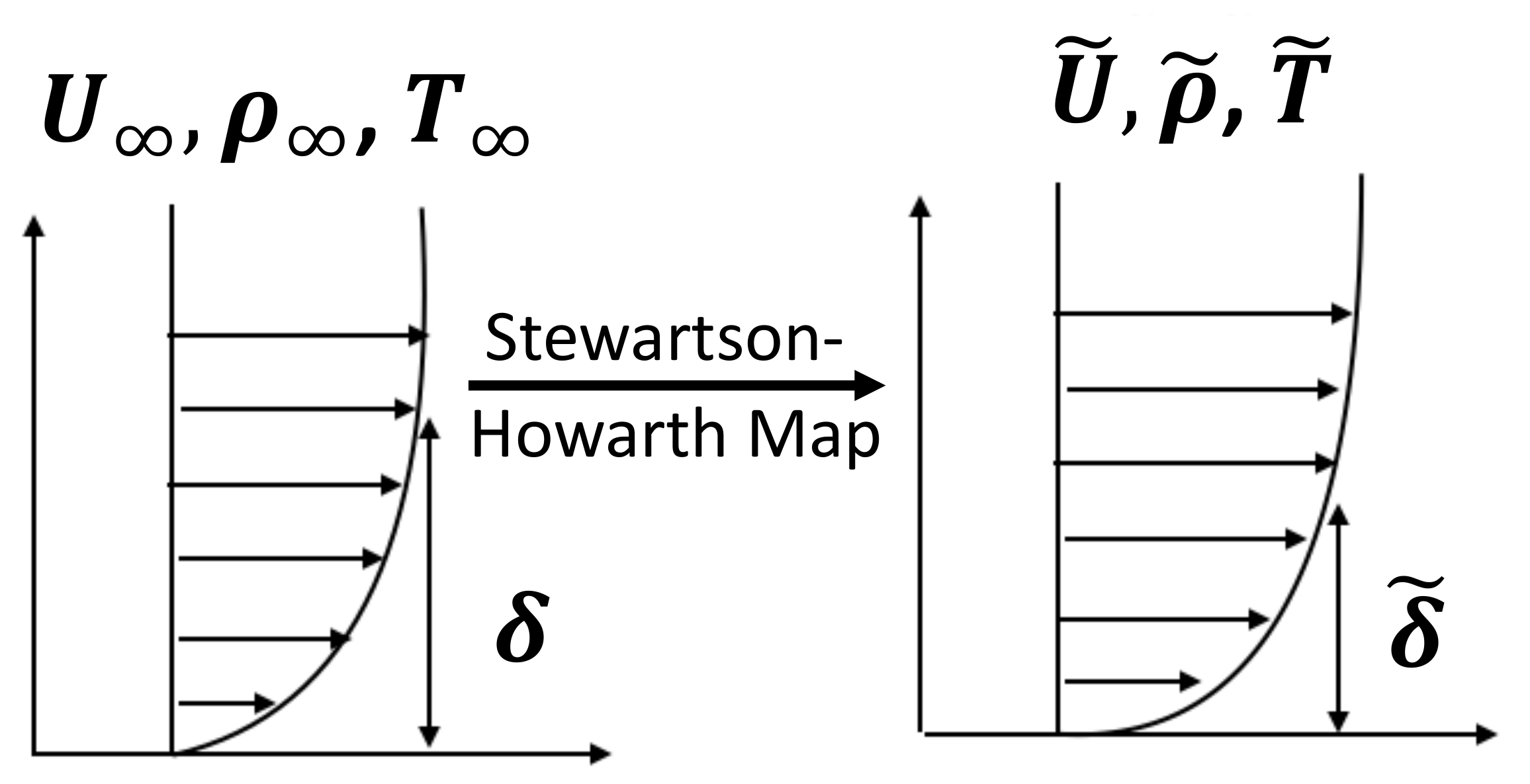}
   \label{compressible_sphere}
   }    
        \subfigure[Compressible flow (supersonic)]
  {
   \includegraphics[width=0.4650\linewidth, angle =0]{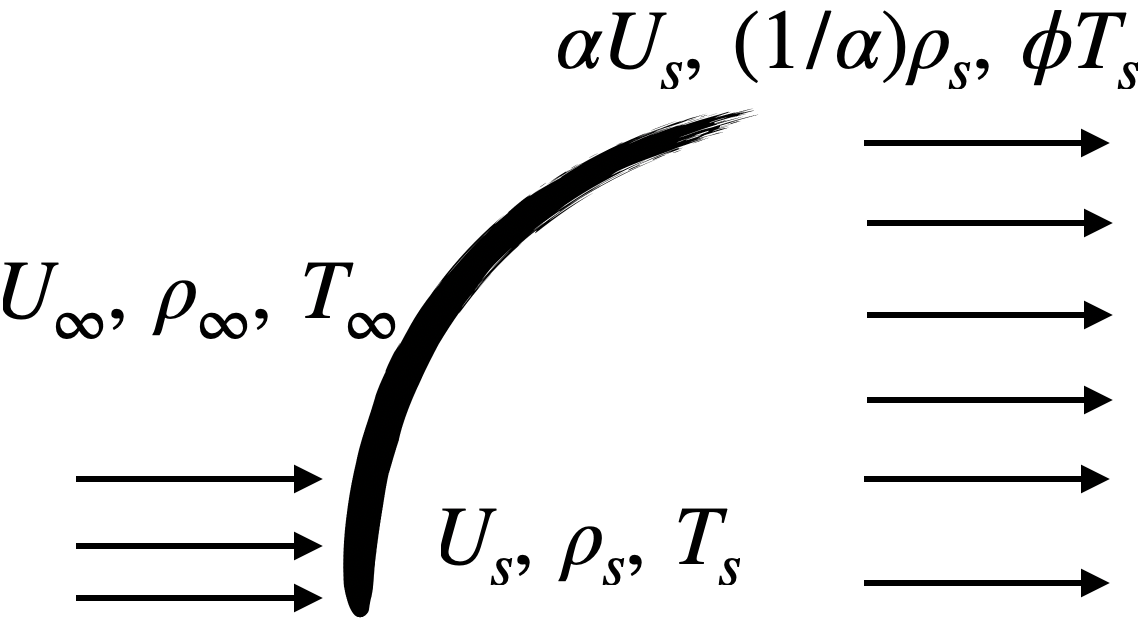}
   \label{shock_wave}
   } 
        \subfigure[Rarefied flow (transitional regime)]
  {
   \includegraphics[width=0.4650\linewidth]{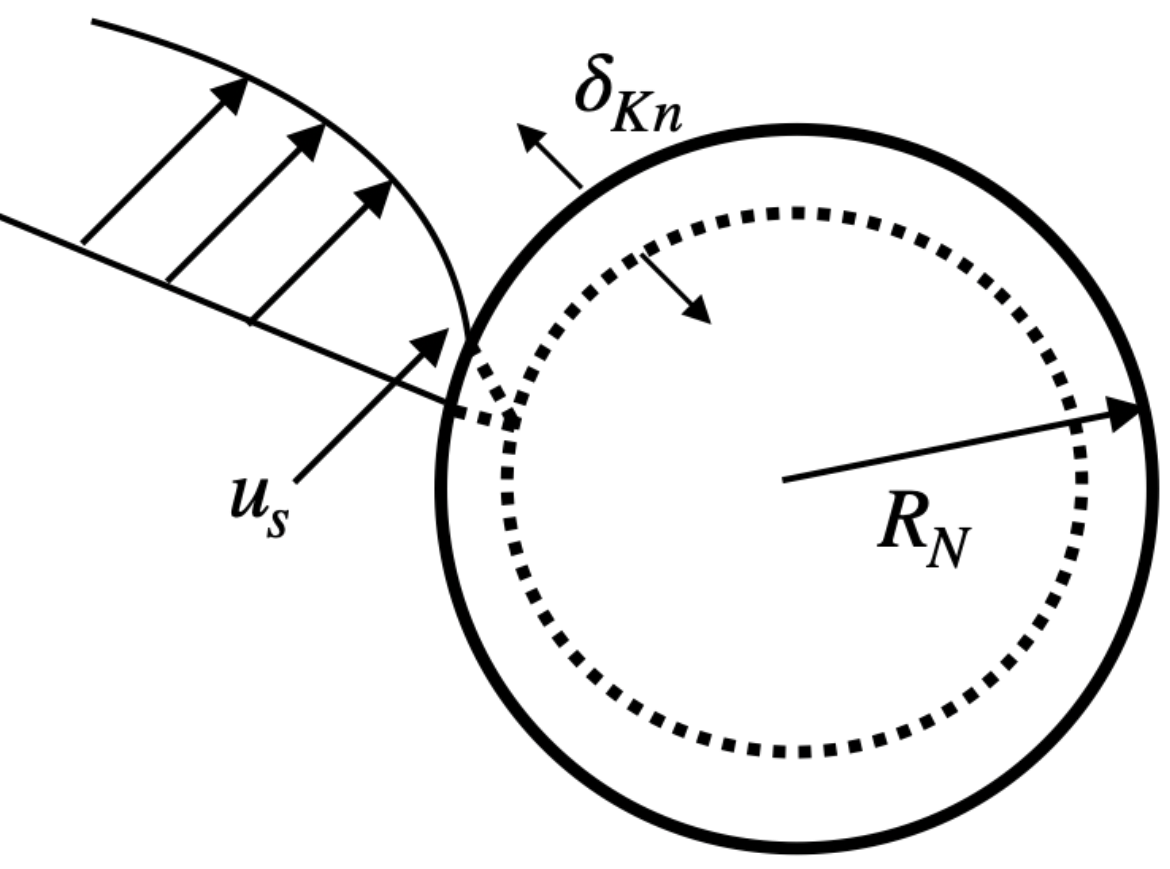}
   \label{rarefied_sphere}
   } 
   \caption{Flow over sphere in different flow regimes depicting correlation development approach}
   \label{sphere_bl_incomp}
 \end{figure}
\textbf{Incompressible flow:} For sufficiently high $Re$ ($\gg 1$), as viscous forces are confined within a thin region called the boundary layer near the wall,
\cite{abraham1970functional} considered drag force on an effective sphere (of radius $R+\delta$), which includes the boundary layer ($\delta$) region as shown in Fig.~\ref{incompressible_sphere}. The drag force on the effective sphere is only a function of free-stream momentum flux, and the effects of viscosity on this extended sphere can be ignored.  
Therefore, on the effective sphere, $C_d$ is equal to a constant ($C_0$), which is independent of $Re_\infty$. The drag force on the actual sphere ($F_d$) can then be expressed as:
\begin{equation}
    F_d = C_0 \left(1+\cfrac{\delta}{R} \right)^2 \cfrac{\rho_{\infty} \pi R^2}{2} U_{\infty}^2
    \label{Force_ic}
\end{equation}
where $F_d$ is drag force and $\delta$ is boundary layer thickness. The dimensional analysis arguments for obtaining the expression for drag force in Eq.~\ref{Force_ic} are also supported by mathematical arguments \citep{imai1957theory} and related discussion can be found in Ref.~\citep{dyke1971comments}. The coefficient of drag ($C_d^{ic}$) from $F_d$ on the original sphere can be expressed as: 
\begin{equation}
    C_d^{\text{ic}}= C_0 \left[1+ \cfrac{\delta_0}{(Re)^{1/2}}\right]^2
    \label{Incompressible_Cd}
\end{equation}
where $\delta/R \approx 
\delta_0/\sqrt{Re}$, the superscript `$ic$' denotes incompressible regime, and $Re_{\infty}=\rho_\infty U_\infty D/\mu$ for diameter $D$ of the sphere, and $C_0 \delta_0^2 = 24$, which reduces to Stokes's result for low $Re$. Abraham \cite{abraham1970functional} found $\delta_0 = 9.06$ based on fitting the correlation to limited experimental data.

\begin{figure}
   \centering
  {
    \includegraphics[width=0.50\linewidth]{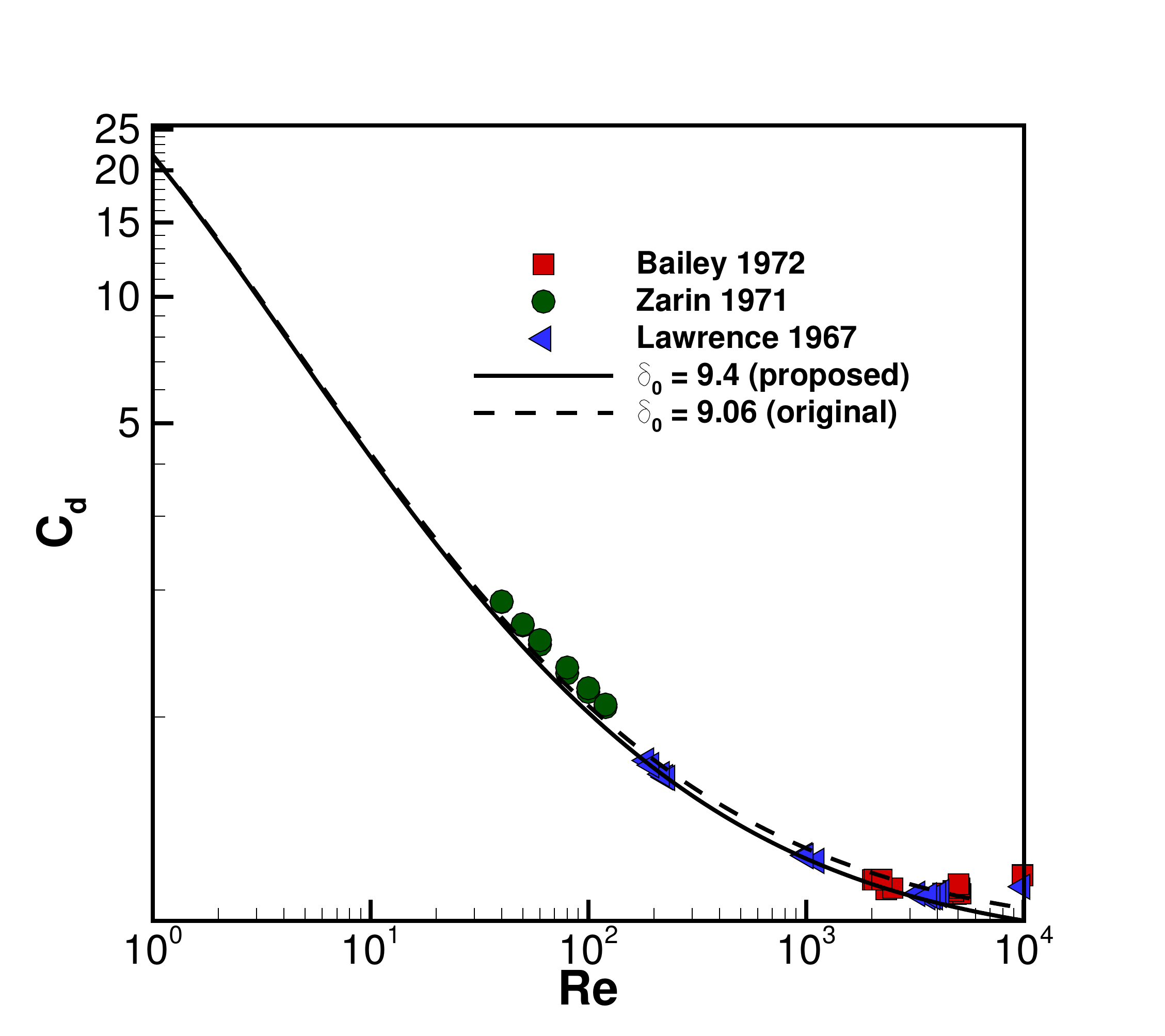}
   }  
   \caption{Comparison of Eq.~\ref{Incompressible_Cd} for drag coefficient in incompressible limit. Experimental data is taken from Refs.~\citep{Bailey_Hiatt}, \citep{Zarin_Nicholls}, and \citep{Lawrencedrag}.}
   \label{compare_delt_o}
 \end{figure}

\textbf{Subsonic compressible flow ($M_{\infty} \leq 1.0 $):}  Compressibility results in density changes, which for $M_{\infty}<1.0$ can be accounted by applying a correction to the incompressible formulation. 
We map the boundary layer of a weakly compressible flow to an equivalent incompressible flow using \cite{howarth1948concerning} and \cite{stewartson1949correlated} transformation. The schematic showing the transformation ($U_{\infty} \rightarrow \tilde{U}, \rho_{\infty} \rightarrow \tilde{\rho},T_{\infty} \rightarrow \tilde{T}$) is presented in Fig.~\ref{compressible_sphere}, and algebraic details are given in the Supplementary Information. Including compressibility effects, the expression for drag coefficient for compressible flow is:
\begin{equation}
\begin{split}
     C_d^{\text{c}}=
     C_0 \ \Theta_{\infty}  \left[1+\cfrac{\delta_0}{(\tilde{Re}_{\infty})^{1/2} }\right]^2
     \end{split}
     \label{drag_comp_1}
\end{equation}
 where
\begin{equation}
\Theta (M) = \left(\cfrac{\tilde{T}}{T}\right)^{\gamma/(\gamma-1)} = \left[ 1+(\gamma-1)\cfrac{M^2 }{2} \right]^{\gamma/(\gamma-1)}
\end{equation}
\begin{equation}
\tilde{Re} (Re, M) = Re  \ \Theta(M)^{\cfrac{\gamma+1}{2\gamma}-\cfrac{\gamma-1}{\gamma}\omega}
\label{tilde_Re}
\end{equation}
where $\tilde{Re}_{\infty}$ and $\Theta_{\infty}$ are respectively equivalent to $\tilde{Re} (...)$ and $\Theta (...)$ evaluated at free-stream conditions ($ Re_{\infty}, M_{\infty}$),  $\gamma$ is the ratio of specific heat capacities at constant pressure and volume and $\omega$ is the exponent in the power-law dependence  of viscosity ($\mu \propto T^\omega$) on the temperature.  
Before we proceed to supersonic flow regime, we point out that accurate prediction in the subsonic regime requires modification of the parameter $\delta_0$ to $9.4$ from $9.06$. The modified $\delta_0$ does not modify incompressible drag, as shown in Fig.~\ref{compare_delt_o}, compared to the original value of $\delta_{0}$.

\textbf{Supersonic Flow ($1 \leq M_\infty \leq 5 $):} 
When a flow becomes supersonic, a thin region known as a shockwave causes rapid compression of the gas. For normal shocks, an analytical solution can be obtained relating the pre and post-shock states of the gas; these relations are known as Rankine-Hugoniot jump conditions . The post-shock state always corresponds to an increase in temperature, density and pressure, and a decrease in flow field velocity (which is always subsonic).  However, for blunt bodies the shock is not normal and curves significantly around the body of interest. This is commonly known as a bow shock. Bow shocks are formed around spherical bodies for supersonic and hypersonic flows.

To extend the drag model to supersonic conditions, the same formulation developed in Eqs.~\ref{drag_comp_1} and \ref{tilde_Re} is employed, with an additional contribution to the drag ($Cd_p$) due to pressure changes across shock. The effect of compression due to the shock-wave is approximated by using a set of effective flow variables. First, the Rankine-Hugoniot conditions across a normal shock are used to obtain post-shock conditions (given in the Supplementary Information) ($U_{\infty} \rightarrow U_s,\rho_{\infty} \rightarrow \rho_s, T_{\infty} \rightarrow T_s, p_{\infty} \rightarrow p_s$). To approximate effects of the non-normal (hyperbolic) structure of the shock,
we introduce mapping functions $\alpha (M_\infty)$ and $\phi (M_\infty)$ to transform post-shock variables ($U_{s} \rightarrow \alpha U_s,\rho_{s} \rightarrow (1/\alpha) \rho_s, T_{s} \rightarrow \phi T_s, p_{s} \rightarrow p_s^{\text{ef}},M_s \rightarrow M_s$) as shown in Fig.~\ref{shock_wave}. The inverse scaling ($\alpha$ and $1/\alpha$) for density, and velocity, respectively ensures mass conservation. The function $\phi$ is obtained by the constraint that the transformation preserves post-shock Mach number ($M_s$), yielding $\phi = \alpha^2$. Using the transformation of the flow variables, the pressure change term ($Cd_p$)  can be approximated as:
\begin{equation}
\begin{split}
   Cd_p =  C_1 \cfrac{1}{(1/2)\rho_{\infty} U_{\infty}^2}\left[ p_{s}^{\text{ef}} -p_{\infty} \right]
\end{split}
\end{equation}  
where $p_{s}^{\text{ef}}$ is the effective pressure and $C_1$ is the scaling for the contribution of pressure change across a curved shock.
The pressure term ($Cd_p$)  does not contain an effective area term,  which includes boundary layer thickness, because the pressure does not significantly vary within the boundary layer and therefore is applied directly on the original sphere of radius ($R$).
Expressing the pressure change in terms of the difference in momentum flux between free-stream (at $ \rho_\infty$ and $U_\infty$) and the effective momentum flux (at $ (1/\alpha) \rho_s$ and $\alpha U_s$)  after the shock-wave, we get
\begin{equation}
\begin{split}
    Cd_p  = \left[ C_1 \left(1-\alpha \cfrac{U_s}{U_{\infty}}\right) \right] 
\end{split}
\end{equation} 
Therefore, the expression for overall drag coefficient becomes:
\begin{equation}
\begin{split}
      C_d^{\text{c}}  = \left[C_1 \left(1-\alpha \cfrac{U_s}{U_{\infty}}\right) \right]+ \left[C_0 \Theta_s \right] \left(1+\cfrac{\delta_0}{(\tilde{Re_s})^{1/2} } \right)^2 
    \label{cd_cont_1}
\end{split}
\end{equation} 
where the second term in Eq.~\ref{cd_cont_1} is the same as in Eq.~\ref{drag_comp_1} but is now evaluated at the effective Mach number, $M_s$, and $\tilde{Re}_s$. $M_s$ is invariant under the proposed transformation, and $\tilde{Re}_s$ is given by:
\begin{equation}
\tilde{Re}_s = Re_{\infty}\left[\cfrac{1}{\alpha^2}\cfrac{T_\infty}{ T_s}\right]^\omega \Theta_s^{\cfrac{\gamma+1}{2\gamma}-\cfrac{\gamma-1}{\gamma}\omega}
\label{tilde_Re_s}
\end{equation}
 where $\Theta_s$ is equivalent to $\Theta$ evaluated at effective conditions after transformation. The temperature ratio raised to the power $\omega$ appears in Eq.~\ref{tilde_Re_s} due to the ratio of viscosities across the shock. While the estimation of the transport coefficients at high-temperature \citep{mankodi2020collision,oblapenko2020influence}, the power-law with $\omega = 0.74$ is the most widely used relation for viscosity at high temperatures for air ($> 600$ K). Furthermore, although the Stewartson-Howarth transformation, utilized to derive the function $\Theta$, uses $\omega =1$ for algebraic simplification, the correlation is not sensitive to the magnitude of $\omega$ in the $\Theta$ function.  The algebraic details for $\tilde{Re}_s$ are provided in the appendix (refer to Sec.~\ref{append:supersonic_regime}). 
 There are two unknown parameters $C_1$ and $\alpha$, which are required in Eq.~\ref{cd_cont_1} to determine $C_d$.  For estimating the expression of $\alpha$, owing to the difficulty of analytically obtaining a solution, we rely on heuristic arguments.
 For $M_\infty \gg 1$, for the effective flow to remain continuum ($\tilde{Kn}_s<0.1$, see Eq.~\ref{kn_def}), $\tilde{Re}_s$ should not reduce to zero, implying $\alpha \propto 1/M_{\infty}$ ($T_\infty/T_s \propto 1/M_\infty^2$ for $M_{\infty} \gg 1$). A simple formulation for $\alpha$ satisfying this condition (in addition to $\alpha$ being unity for $M_{\infty} = 1$) is:
\begin{equation}
   \alpha = \cfrac{1}{\alpha_0 \ M_\infty+1-\alpha_0}
\end{equation}
where $\alpha_0$ is an unknown constant parameter. Recall $C_1$ is a scaling factor for the pressure term and is obtained from the solution for the hypersonic limit in the next subsection.

\textbf{Hypersonic Limit ($M_\infty \geq 5 $):}
 In hypersonic flow, the post-shock gas is in thermal and chemical nonequilibrium, and the simplified assumption of calorically perfect gas utilized in supersonic flows become inaccurate. In fact, nonequilibrium reaction chemistry modeling, calculation of transport coefficients, and development of appropriate boundary conditions for hypersonic flows are an active area of research. The objective of the present work is to find a scaling of the drag coefficient in the hypersonic regime, an approximation to which has been obtained by \cite{hornung2019hypersonic}  under the assumption of negligible viscous effects. Hornung \textit{et. al} show that the magnitude of the drag coefficient approaches a nearly constant value at high $M_{\infty}$, which is also supported by earlier theoretical investigations \citep{lighthill1957dynamics} and experimental data.
In current work, we utilize the hypersonic limit ($C_d^{M_\infty} \approx 0.9$, please refer to Fig.~17 in Ref.~\citep{hornung2019hypersonic} for more details) to estimate the expression for $C_1$ in the proposed formulation (Eq.~\ref{cd_cont_1}) for the drag correlation. Substitution of variables ($U_s/U_{\infty}, M_s^2, \Theta$, and $\alpha$) in the limit of $M_{\infty} \gg 1$ in Eq.~\ref{cd_cont_1} and equating the corresponding $C_d$ to $C_d^{M_\infty}$ yields the expression for $C_1$:
 \begin{equation}
\begin{split}
   C_{1} = \cfrac{C_d^{M_\infty}-C_{0}\Big[1+\cfrac{(\gamma-1)^2}{4\gamma}\Big]^{\gamma/(\gamma-1)}}{1-\cfrac{1}{\alpha_0 \ M_\infty}\cfrac{\gamma-1}{\gamma+1}}
\end{split}
\end{equation}
For more algebraic details or the exact high Mach number limit of the variables, please refer to Sec.~\ref{append:hypersonic} of the appendix. At this stage our continuum formulation for the expression of the drag coefficient is complete (Eq.~\ref{cd_cont_1}), with only one unknown parameter $\alpha_0$. Before, we estimate $\alpha_0$, the extension of the model to rarefied regime is presented next. 
 
\subsection{Theoretical Development: Rarefaction Regime}
In rarefied flows ($Kn_{\infty} > 0.01$), due to fewer gas molecule collisions near the surface, the bulk velocity and temperature of the gas do not equilibrate with the surface velocity and surface temperature, respectively. This results in a finite velocity slip and temperature jump at the wall.
To account for these non-continuum effects, velocity slip (and temperature jump) at the wall is employed along with the Navier-Stokes equations. At higher $Kn_{\infty}$ ($> 0.5$), the stress tensor (and heat flux vector) does not depend linearly on the velocity gradient (and temperature gradient); therefore, the Navier-Stokes equations become inaccurate in high $Kn_{\infty}$ regimes. If $ M_{\infty} $ is also high, as molecules travel larger distances without collisions at high $Kn_{\infty}$, the shock-layer and boundary layer merge with each other. Although the Boltzmann equation is applicable for arbitrary degree of rarefaction ($\forall Kn_{\infty}$), the equation is computationally expensive to solve. However, approximate scaling laws relevant to drag correlation can still be informed from the Boltzmann equation, which is adequate for the current work and is described in the next subsections.

\textbf{Rarefaction correction ($\forall Kn_{\infty}$) at low $M_{\infty}$:}  For low speed flows, the Boltzmann equation has been theoretically solved in an approximate manner by \cite{phillips1975drag} for estimating the drag force on a sphere.  Phillips  expressed $C_{d} = f_{Kn} C_{d}^{c}$, where  $f_{Kn}$ is a multiplicative rarefaction correction  to the continuum drag coefficient ($C_d^c$). For low speed flows, numerically identical to the expression by Phillips \citep{phillips1975drag}, an alternate simple closed form expression for $f_{Kn}$ has been proposed \citep{clift1978bubbles} (and for non-spherical particles in Ref.~\citep{dahneke1973slip}) as follows:
\begin{equation}
    f_{Kn} = \cfrac{1}{1+Kn_{\infty}\left[A_1+A_2\exp[-A_3/Kn_{\infty}]\right]}
    \label{rarefied_correction}
\end{equation}
where $A_1 = 2.514 , A_2 = 0.8 ,$ and $A_3 = 0.55 $.
The correction in Eq.~\ref{rarefied_correction} was earlier proposed by  \cite{millikan1923coefficients} and is valid for all Knudsen numbers \citep{clift1978bubbles} for low $M$ numbers. Basically,  $f_{Kn}$ reduces to the high-Kn number drag expression for high $Kn (\gg 1)$ and unity at low $Kn_{\infty} (<0.01)$. An intuitive argument for the justification of Eq.~\ref{rarefied_correction} using a reduced sphere due to slip effects is presented in the appendix (Sec.~\ref{append:rarefaction}). Next, we develop the correction function for high $Kn_{\infty}$ and high $M_{\infty}$.

\textbf{Rarefaction correction for the slip  ($ 0.01 \leq Kn_{\infty} < 0.1$) and transition ($ 0.1 \leq Kn_{\infty} < 10$) regime at high $M_{\infty}$ :} For high Mach number flows ($M_{\infty}>0.3$) flows, an analytical solution of the Boltzmann equation is challenging, and in many cases, impossible.
Several sets of non-linear constitutive relations and higher order moment equations have been proposed \citep{singh2016onsager,singh2017derivation,struchtrup2005macroscopic} as computationally efficient alternatives to the Boltzmann equation for modeling slip and transition regimes. For high-speed flow over a sphere, Singh and Schwartzentruber \citep{singh2016jfm1,singh2017jfm2} mathematically showed that the ratio of the non-linear to linear (used in the Navier-Stokes equations) constitutive relations dominantly depends on a non-dimensional number called $W_r^T$, which is given by: 
\begin{equation}
    W_r^T = W_r \left(1+\cfrac{T_p}{T_s}\right)^\omega
    \hspace{0.5in} M_{\infty} > 1
    \label{Wrt}
\end{equation}
where

\begin{equation}
   W_r = \cfrac{M_{\infty}^{2\omega}}{Re_{\infty}} = Kn_{\infty} \ M_{\infty}^{2\omega-1}\sqrt{\cfrac{2}{\pi \gamma}}
 \end{equation}
and $T_p$ is surface temperature. In fact, an approximate contribution of the higher order (up to infinite order in terms of $Kn_{\infty}$) constitutive relations has been shown as a scaling factor to the heat transfer coefficient. At high $Kn_{\infty}$, the gas does not fully thermally accommodate with the wall, and the drag force depends explicitly on the surface temperature as also evident in Eq.~\ref{Wrt}. 
We employ the same correction to $f_{Kn}$, which modifies the high Knudsen number correction using an additional term based on $W_r$:
 \begin{equation}
 \begin{split}
    f_{Kn, W_r} = \cfrac{1}{1+Kn_{\infty}\left[A_1+A_2\exp(-A_3/Kn_{\infty})\right]}
    \cfrac{1}{1+\alpha_{hoc} W_r^T}
    \end{split}
    \label{rarefied_correction_2}
\end{equation}
Equation~\ref{rarefied_correction_2} corrects the drag coefficient for rarefaction effects at moderate $Kn_{\infty}$
and at high $M_\infty$. 

\textbf{Rarefaction correction for free-molecular regime ($Kn_{\infty}\geq10$) at high $M_{\infty}$:}  As a final step, we bridge the correlation to the analytically obtained $C_d$ for the free-molecular regime. An analytical expression for $C_d$ valid for free-molecular flow has been developed by \citep{patterson1971introduction} and is given by:
\begin{equation}
\begin{split}
   Cd_{fm} = \cfrac{(1+2s^2) \text{exp}(-s^2)}{s^3\sqrt{\pi}}+\cfrac{(4s^4+4s^2-1)\text{erf}(s)}{2s^4}+\cfrac{2}{3s}\sqrt{\pi \cfrac{T_p}{T_\infty}}
   \end{split}
   \label{cdfm}
\end{equation}
where $s = M_{\infty} \sqrt{\gamma/2}$.  \cite{singh2017jfm2} proposed a bridging function between a purely empirical drag correlation for high $M$($>1$) transition and free-molecular regimes using the inverse Cheng's parameter \citep{cheng1961hypersonic}, $K_c$. Mathematically, $1/K_c^2 \propto \mu^{*} T_\infty/\mu_{\infty} T^*$, where $T^*$ is the average of $T_s$ and $T_p$ and $\mu^{*}$ is evaluated at $T^*$. Substituting power-law relations for viscosity, it can be shown that the inverse of $K_c$ is proportional to $W_r^T$, both of which depend on the wall temperature that becomes important as the degree of rarefaction increases.  In this work, we employ $Br (\propto W_r^T)$ as the correlation parameter to bridge the proposed general physics-based drag correlation to the free-molecular expression,
\begin{equation}
\begin{split}
   Br = W_r^T\cfrac{M_{\infty}^{2\omega-1}+1}{M_{\infty}^{2\omega-1}}
    \end{split}
\end{equation}
 $Br$ has the desirable properties that for $M_{\infty} \gg 1 \implies Br \rightarrow W_r^{T}$ and  for $M_{\infty} \ll 1 \implies Br \rightarrow Kn_{\infty}$.
Using a rational polynomial function as a plausible bridging function, the expression for full drag correlation is defined as:
\begin{equation}
    C_d = C_d^{\text{c}} (M, Re) f_{Kn,W_r}\cfrac{1}{1+Br^\eta} + Cd_{fm}\cfrac{Br^\eta}{1+Br^\eta}
    \label{final_drag_correlation}
\end{equation}
 where $\eta$ is an unknown parameter. Equation \ref{final_drag_correlation} is the main result of the present work. There are three unknown parameters $\alpha_{0},  \alpha_{hoc} $ and $\eta$, which are obtained in the next subsection.
The complete set of equations, for the entire correlation, are also provided separately in Sec.~\ref{app:proposedmodel} of the appendix.
\subsection{Estimation of $\alpha_0$, $\alpha_{hoc}$, and $\eta$}
 Least-square fitting was used to determine the unknown parameters based on available experimental data and relevant simulations. A summary of the data used for fitting unknown parameters is compiled in this section.
 \cite{Bailey_Hiatt} determined drag on spherical bodies for a wide range of Mach and Reynolds numbers ($0.1 < M_{\infty} < 6$ and $10^1 < Re_{\infty} < 10^5$)  in a ballistic range.  \cite{Bailey_high_Re} also determined the drag on spheres for higher Reynolds numbers using a similar setup. \cite{Sreekanth} determined the drag on spherical bodies in the transitional Knudsen number regime ($M_{\infty} = 2$ and $ 0.1 <Kn_{\infty}<0.8$) using a wind tunnel setup. \cite{Zarin_Nicholls} determined the drag on spherical bodies in a subsonic wind tunnel using magnetic suspension for a range of Reynolds numbers ($0.1 < M_{\infty} < 0.57$ and $4\times10^1 < Re_{\infty} < 5\times 10^4$).  used CFD and DSMC, respectively, to determine the drag on a spherical body \cite{Kissel,Overell,Macrossan}. Aroesty measured spherical drag in Berkeley's low-density wind tunnel for Mach numbers of roughly 2, 4 and 6 for $10^1 < Re_{\infty} < 10^4$ \cite{Aroestydrag}. \cite{Lawrencedrag} used a ballistic range to determine drag for subsonic spheres ($0.1 < M_{\infty} < 1$) for a range of Reynolds numbers ($2\times10^1 < Re_{\infty} <  10^4$). \cite{Charters_drag} also used a ballistic range setup and determined drag on spheres for a larger range of Mach numbers at relatively high Reynolds numbers ($0.3 < M_{\infty} < 4$ and $9.3\times10^4 < Re_{\infty} < 1.3 \times 10^6$).  \cite{Hodges_drag} determined drag on supersonic and hypersonic spheres ($2 < M_{\infty} < 10$) for very high Reynolds numbers ($Re_{\infty} > 3 \times 10^6$). May also found the drag on spherical bodies, but for lower Reynolds numbers and Mach numbers ($0.8 < M_{\infty} < 4.7$ and $1.1\times10^3 < Re_{\infty} <  8.4 \times 10^5$). The obtained parameters from least-square fitting are listed in Table~\ref{vasilevskii_params}.

\begin{table}
\caption{ Parameters required in the proposed correlation }
\begin{tabular}{ccc}
Parameters & Physical Significance  & Magnitude \\
\hline
$\delta_0$ & boundary layer thickness scaling & 9.4 \\
$\alpha_0$& shock-curvature parameter & 0.356 \\
$A_1, A_2, A_3$& low-speed rarefaction correction \citep{clift1978bubbles} & 2.514, 0.8, 0.55 \\
$\eta$& bridging function modulator & 1.8 \\
$\alpha_{hoc}$ & high speed rarefaction correction & 1.27 \\
\end{tabular}
\label{vasilevskii_params}
\end{table}

  \begin{figure}
   \subfigure[]{
    \includegraphics[width=0.45\linewidth]{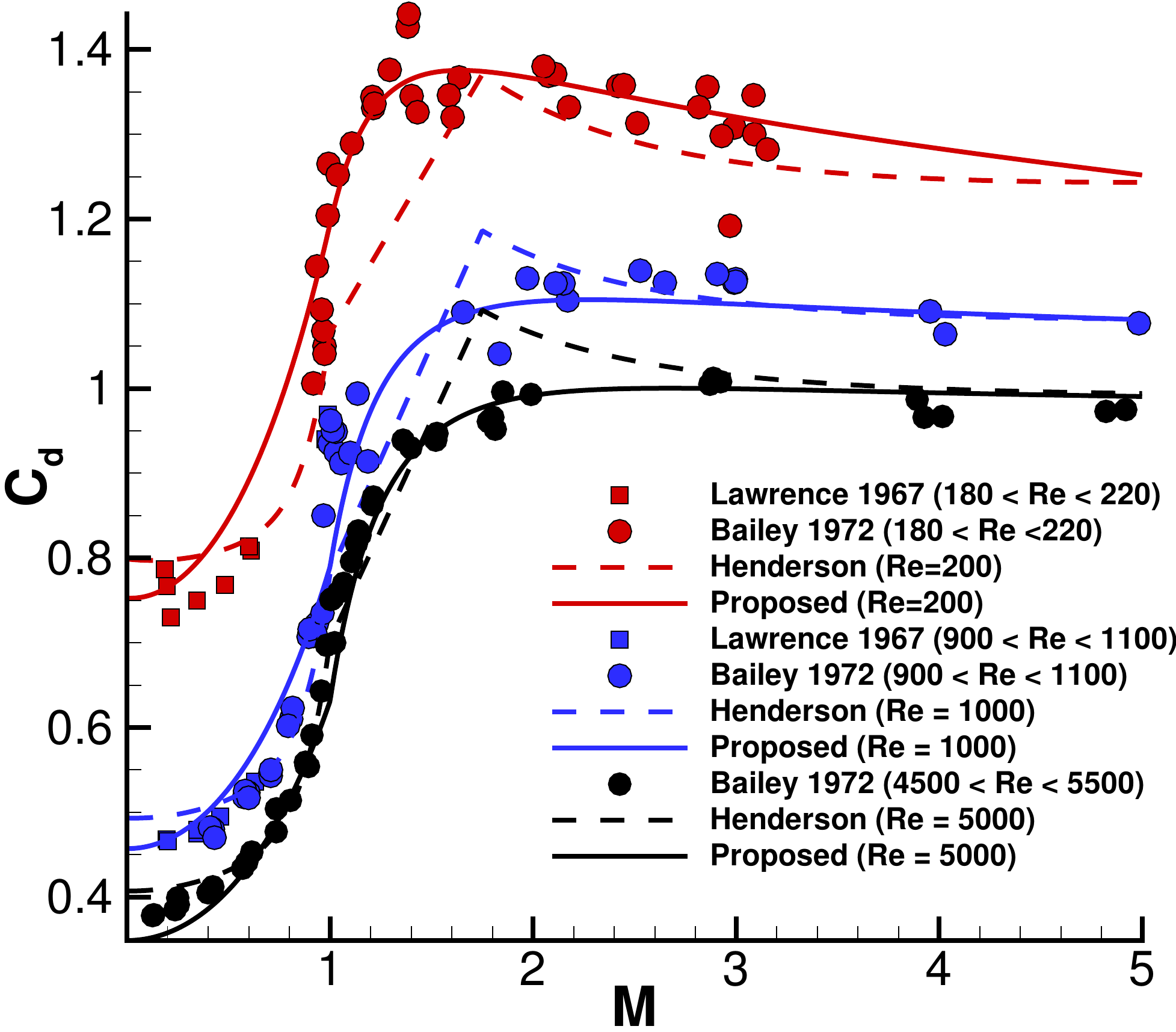}
    \label{cont_Hend_compare}
    }
    \subfigure[]{
     \includegraphics[width=0.45\linewidth]{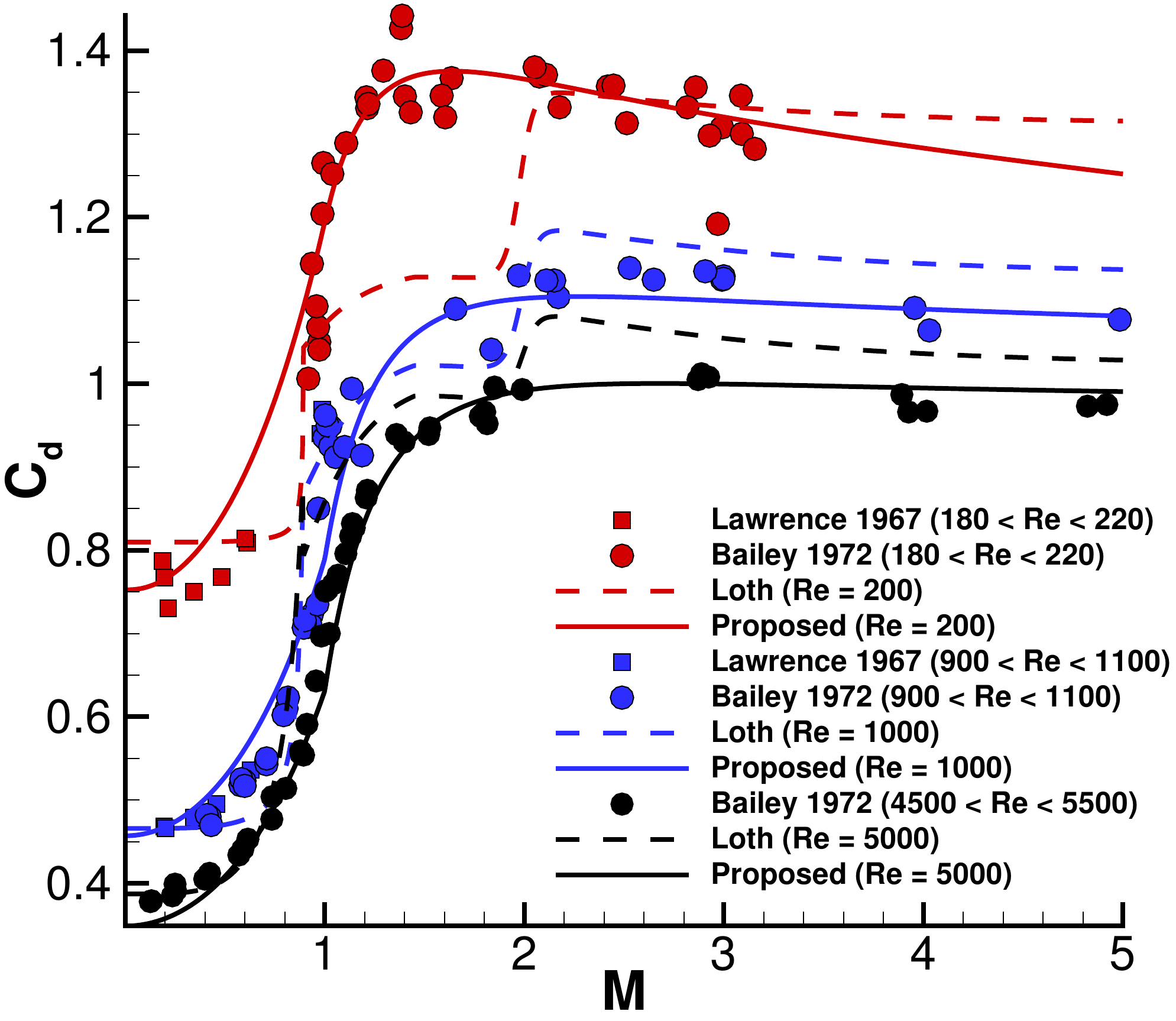}
    \label{cont_Loth_compare}
   }  
   \caption{Comparison of the proposed model with experimental data, Loth model, and Henderson model}
   \label{cont_Hend_Loth_compare}
 \end{figure}

\section{Results and Discussion}
In this section, we compare the proposed drag model with state-of-the-art models. We then present the drag correlation results for different species type (monatomic, diatomic, and triatomic gases). Lastly, we apply the proposed model to investigate sensitivity to surface heating rates for dusty flow over a sphere, relevant for high-speed flows in Mars atmosphere.

\subsection{Comparison to State-of-the-art Models}
We compare the proposed model to standard empirical models that are widely used in the literature. Specifically, the results are compared with \cite{Henderson} and  \cite{loth2008compressibility} models. The equations for both models are reported in in Sec.~\ref{Henderson_Loth_Equation} of the appendix.

In Figure \ref{cont_Hend_compare}, the proposed model is compared with Henderson model and experimental data with Mach numbers for a range of $Re$ corresponding to the continuum regime. For subsonic and hypersonic regimes, both the Henderson and proposed models fit the data adequately. However, near the transonic and supersonic regimes, the proposed model predicts the experimental data more accurately than the Henderson model. This is not surprising; the Henderson model interpolates the drag in these regimes ($1 \geq M_{\infty} \geq 1.75$), while the proposed model has systematically incorporated shock-wave physics. Next, we compare the proposed model with the Loth model.

\begin{figure}
   \centering
  {
    \includegraphics[width=0.5\linewidth]{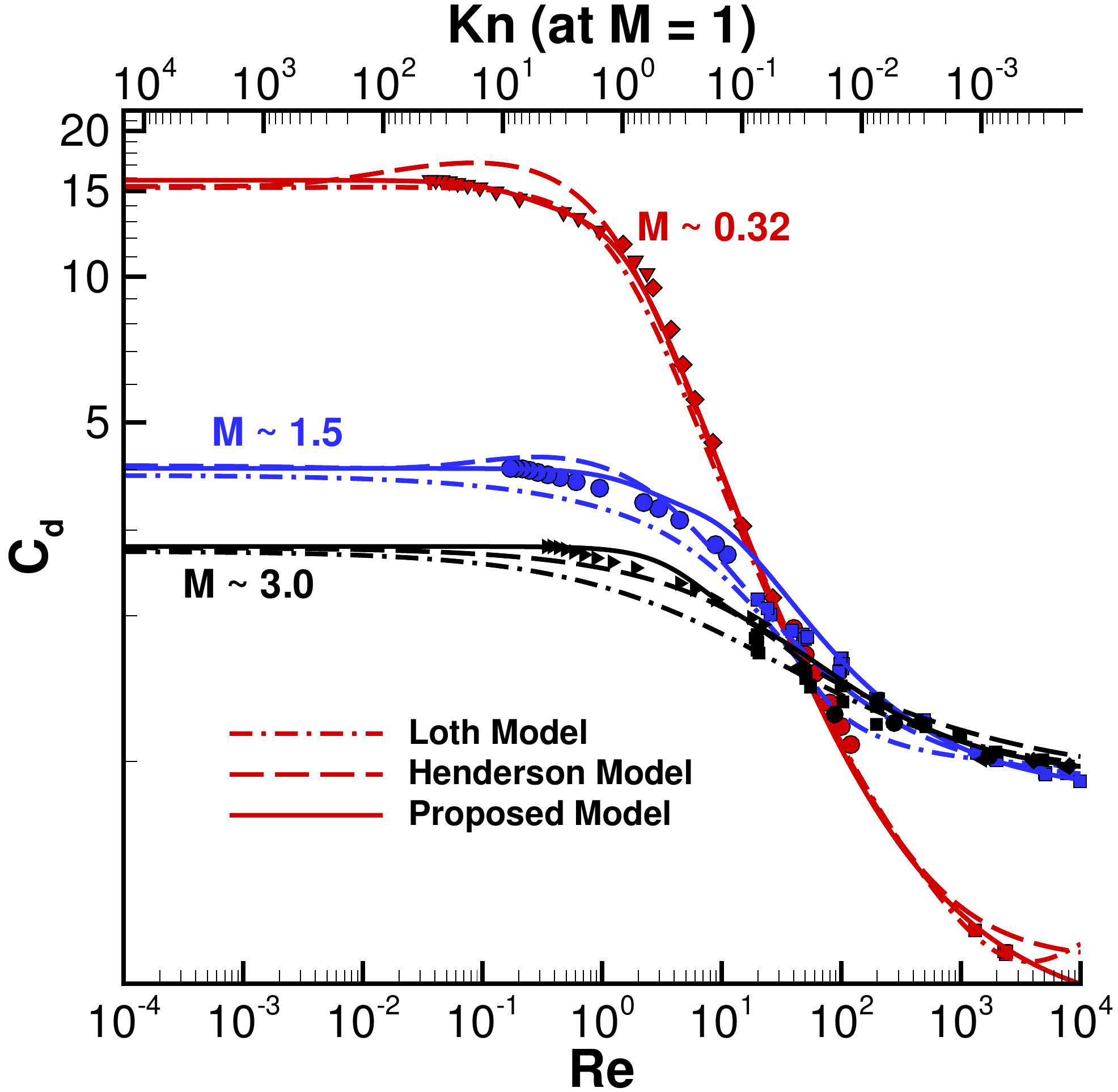}
   }  
   \caption{Comparison of the proposed model, Hendersons model, Loths model, and experimental data  vs $Re$ for $M=0.32, 1.5$ and $3.0$ in red, blue, and black color respectively. Experimental data is taken from several sources:  the red circular symbols from Ref.~\citep{Zarin_Nicholls}, red squares from Ref.~\citep{Lawrencedrag}, blue and black squares from Ref.~\citep{Bailey_Hiatt}, and black diamonds from Ref.~\citep{May_drag}. The red triangles, red diamonds, blue circles, and black triangles correspond to the DSMC data taken from Ref.~\citep{li2019mass}.  The top axis denotes Knudsen numbers, evaluated at $M_{\infty} = 1$ using Eq.~\ref{kn_def}. }
   \label{noncont_Re_compare}
 \end{figure}

  \begin{figure}
    \subfigure[]{
    \includegraphics[width=0.30\linewidth,trim={0.0cm 0.0cm 0.0cm 0.00cm}]{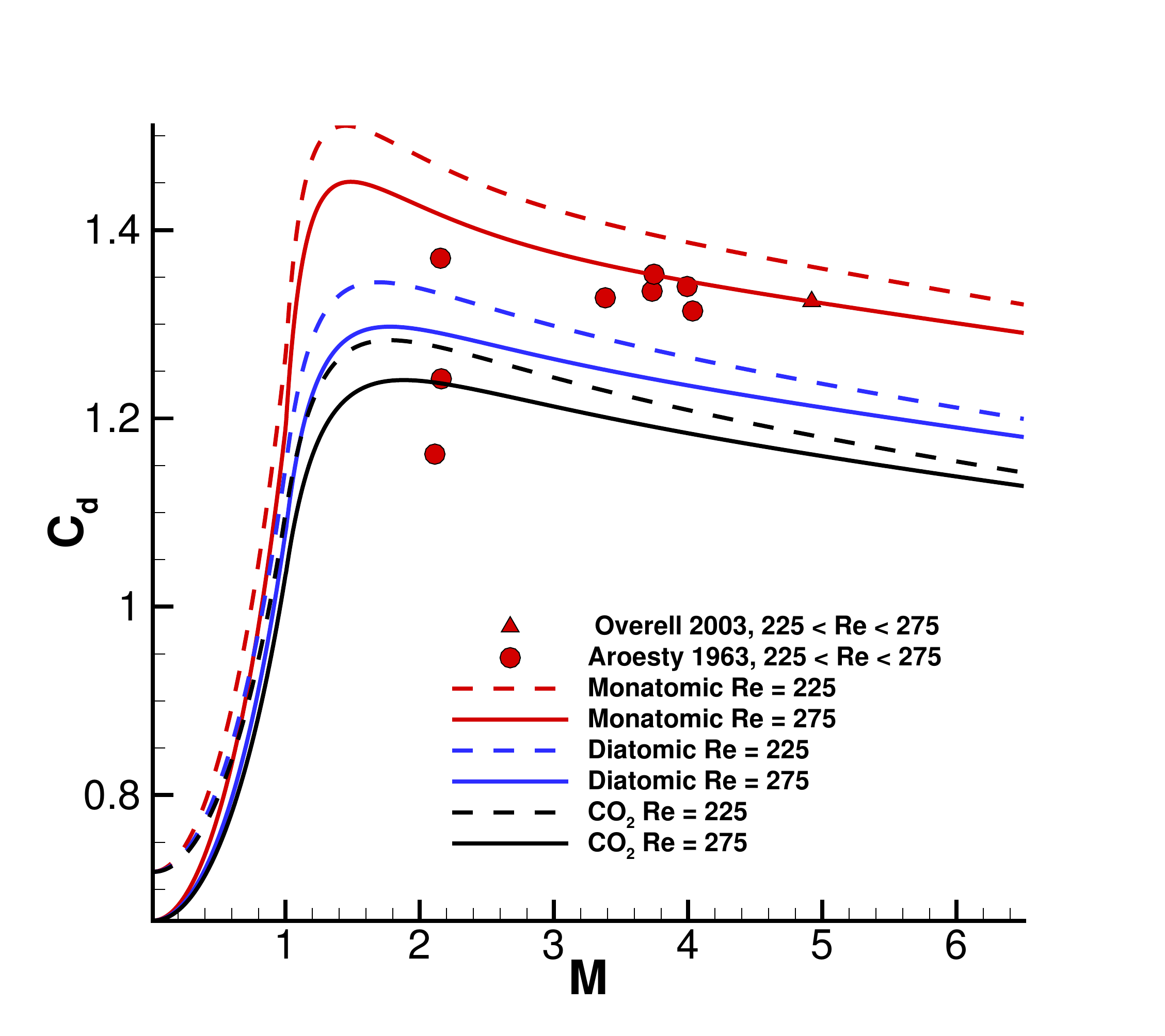}
    \label{Cd_gamma_Re_250}
    }
    \subfigure[]{
     \includegraphics[width=0.30\linewidth,trim={0.0cm 0.0cm 0.0cm 0.00cm}]{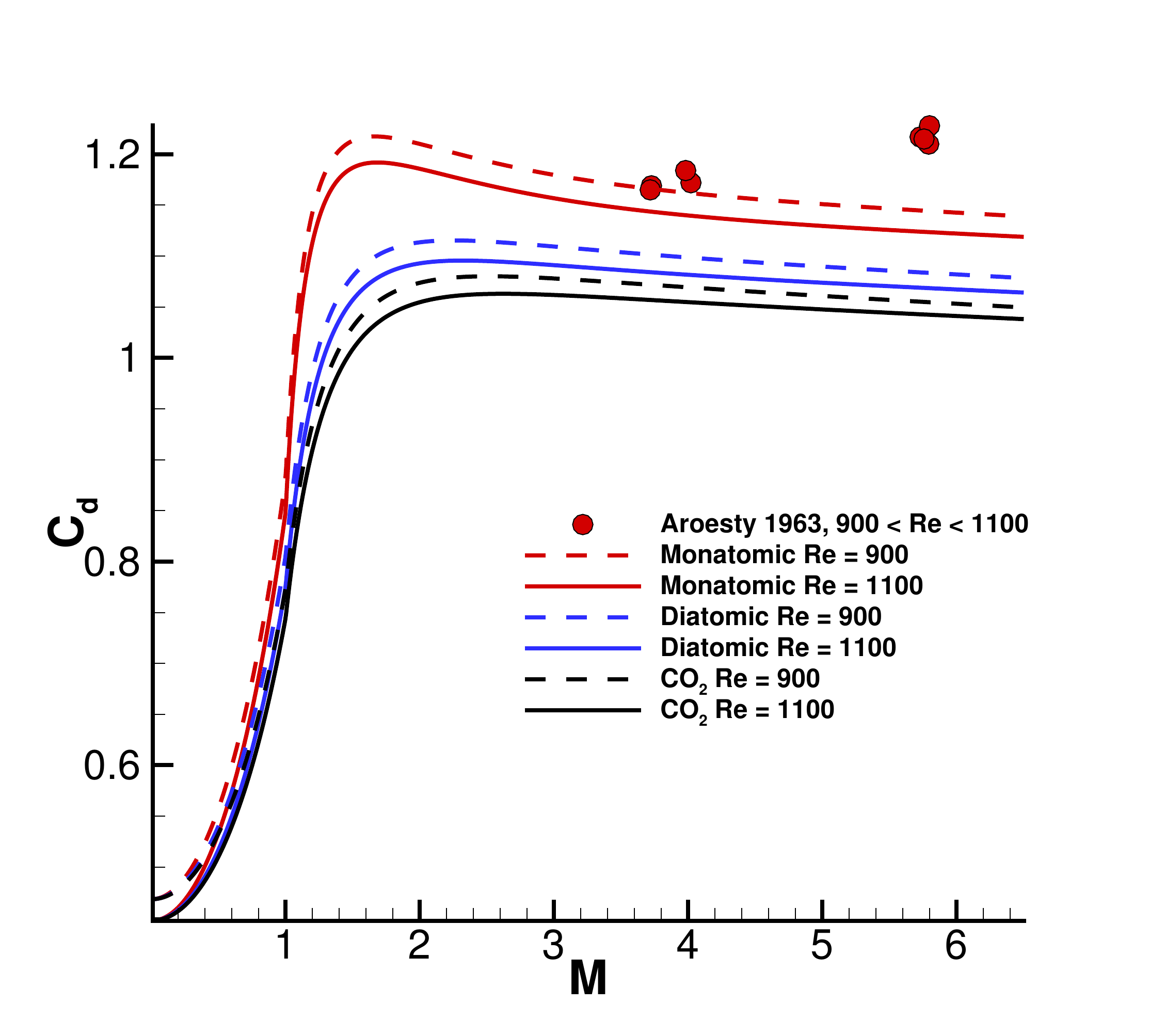}
    \label{Cd_gamma_Re_4000}
   }  
     \subfigure[]{
     \includegraphics[width=0.30\linewidth,trim={0.0cm 0.0cm 0.0cm 0.00cm}]{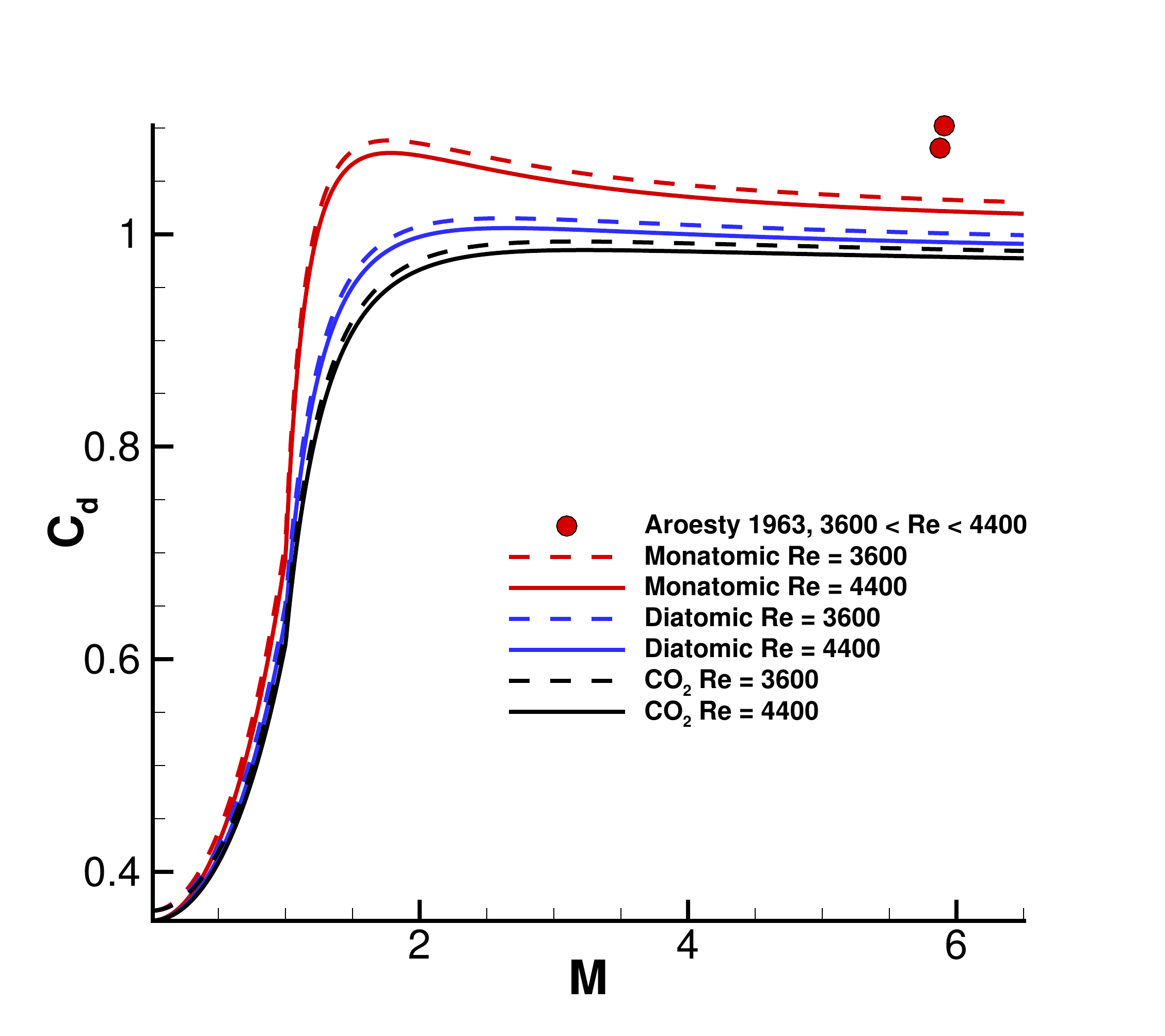}
    \label{Cd_gamma_Re_4000}
   } 
   \caption{Comparison of the proposed model using different gas types. Experimental data from Aroesty and CFD simulations from Overell are also included, both of which considered monatomic gases. }
   \label{gamma_re}
 \end{figure}
 
Figure~\ref{cont_Loth_compare} compares the proposed model with the experimental data and the Loth model for a wide range of $Re_{\infty}$ as a function of $M_{\infty}$ in the continuum regime. While the Loth model predicts the subsonic regime data as accurately as the proposed model, the disagreement at high $M_{\infty}$ numbers is significant, specifically near transonic and supersonic regimes. The Loth model is based on fitting the data such that $C_d$ remains invariant if $M_{\infty}$ is varied but $Re$ is kept fixed at 45. The proposed model does significantly better compared to the Loth model at all $M_{\infty}$. In terms of the entire data considered, the relative $L_2$ norm of the errors of the proposed correlation, Henderson model and Loth model, when compared to the experimental data are $7.6$, $15.8$ and $12.1$ respectively.

Next, we compare the proposed model, Henderson model, and Loth model for a range of $Re_{\infty}$ in Fig.~\ref{noncont_Re_compare}, which includes data from continuum to free-molecular regime for three different free-stream $M_{\infty}$. All three models compare reasonably well with the data with discrepancies in certain regimes. Firstly, the proposed model predicts accurate drag coefficients in the free-molecular (low $Re$) limit for subsonic data supported by the DSMC data at these conditions. However, in the limit of very high $Kn$, the Knudsen number correction from Eq. \ref{rarefied_correction} scales correctly compared to the free-molecular limit, but differs by a constant factor. Loth's model approaches this limit as opposed to the free-molecular limit, which is the reason for the discrepancy for the subsonic data. While the proposed model agrees well with the experimental data, it deviates slightly in the transitional regime. The proposed model under and over-predicts over short a range of $Re$. It is noted that the bridging function is semi-empirical and is based on non-linear constitutive relations scaling and data fitting. Additional data could help improve the parameterization of this transition region.  For instance, the proposed correlation can be used as an input feature for a neural-network approach in the transition regime, and the fidelity could be improved \citep{li2019mass}.

\subsection{Sensitivity of the drag correlation to $\gamma$ (the ratio of specific heat capacities)}
We proceed by examining effects of thermodynamic properties of the gas on the drag coefficient. This subsection demonstrates that since the proposed model is physics-based, the model embeds explicit dependence on the gas-specific properties. For instance, the proposed drag model explicitly depends on the ratio of specific heat capacities at constant pressure and volume, which is different for monatomic, diatomic, and triatomic gas. In Fig.~\ref{gamma_re}, the proposed correlation is plotted against $M$ for three different $Re$ ranges, for monatomic, diatomic, and triatomic linear molecules (CO$_2$ in the present work). Firstly, the drag coefficient is higher for monatomic gas than triatomic gas. This difference is only significant at high $M_{\infty}\geq1$. Since monatomic gas have higher $\gamma$ compared to triatomic gas, they have fewer energy modes for thermal energy re-distribution generated due to the conversion of bulk energy across a shockwave. Therefore, the resulting higher post-shock pressure and velocity for monatomic gases (please refer to Eqs.~\ref{u_post_shock} and \ref{pressure_post_shock}) results in higher drag coefficient. The experimental data from \cite{Aroestydrag} corresponds to monatomic gas (argon), and therefore provides a test-bed to investigate the sensitivity of the drag to $\gamma$.  Figure~\ref{Cd_gamma_Re_250} shows that the proposed correlation for monatomic gas is in reasonably good agreement with experimental data. Alternatively, changing gas-type from triatomic (CO$_2$) to a monatomic gas shifts the predictions from the correlation in the direction of experimental data, for all three ranges of $Re$ considered.
\begin{figure}
    \subfigure[]{
    \includegraphics[width=0.35\linewidth,trim={0.0cm 0.0cm 0.0cm 0.00cm}]{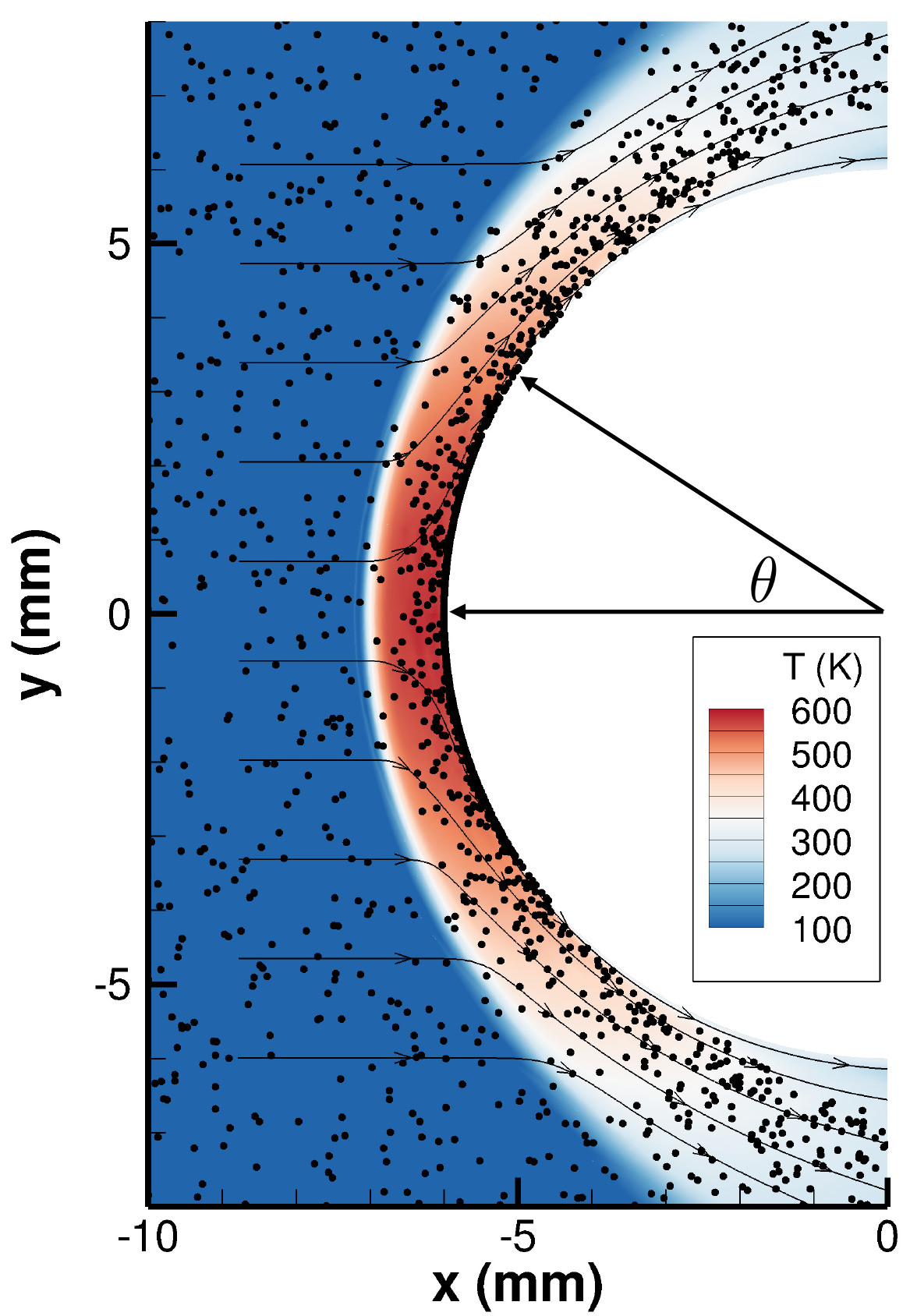}
    \label{temp_eric}
    }
    \subfigure[]{
    \includegraphics[width=0.5\linewidth,trim={0.0cm 0.0cm 0.0cm 0.00cm}]{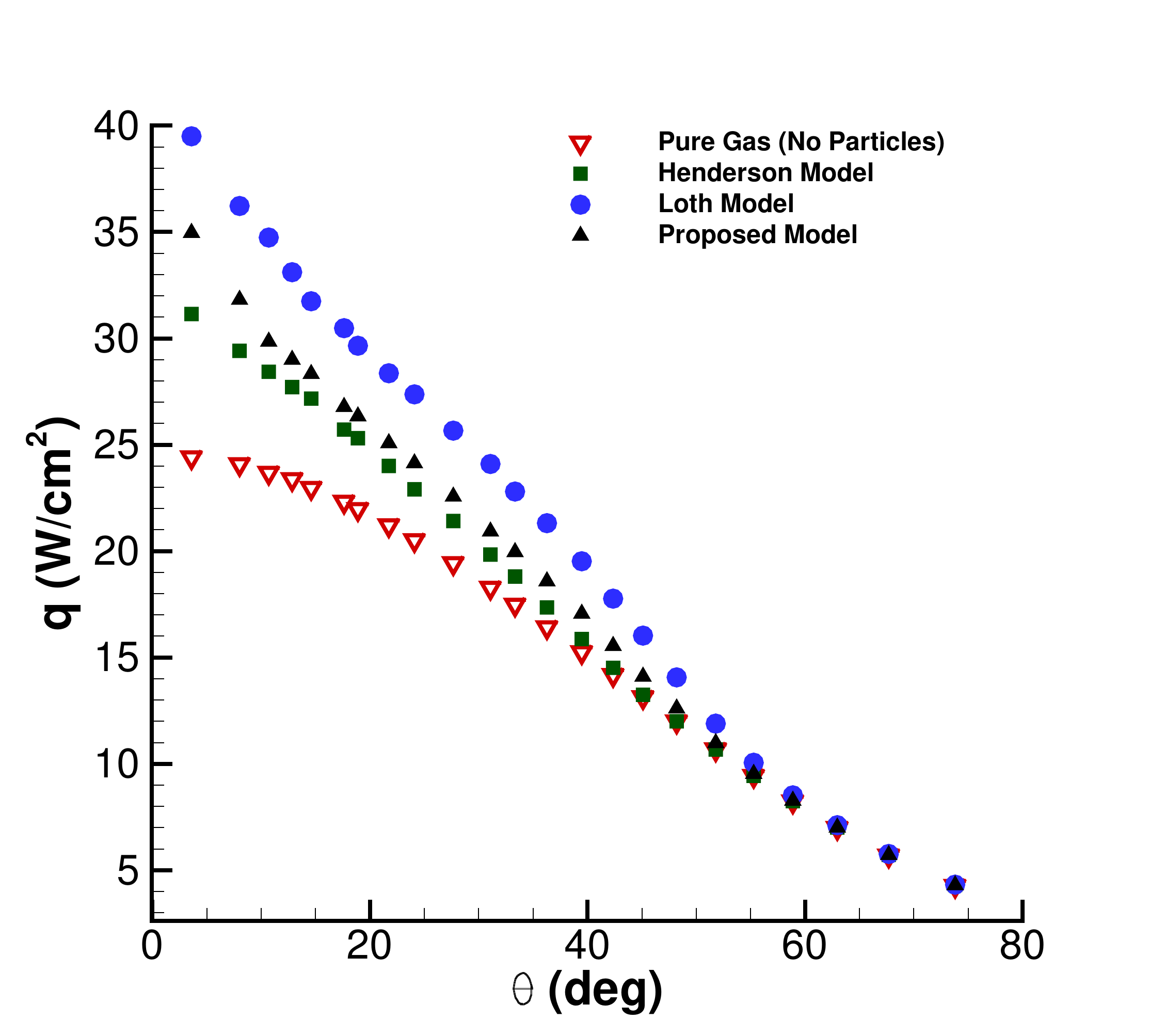}
    \label{heat_flux_eric}
    }
	\caption{Pure-gas and dusty-gas surface heat flux profiles for a flow over a sphere of diameter 0.012 m and surface temperature of 300 K. The free-stream conditions are characterized by $Ma_\infty = 6.1 $, and $P_{\infty} = 1000 $ Pa, $T_{\infty} = 68$ K. The dust particles are made of SiO$_2$, with a density of $2264$ kg/m$^3$, mean diameter of $0.19$ $\mu$m and $3\%$ mass loading ratio (the mass loading ratio is the ratio of the mass flux of the particles to the mass flux of the gas).
	}
	\label{vasilevskii_heating}
\end{figure}
\subsection{Sensitivity of the drag correlation to surface heat flux for a high-speed dusty gas flow over a sphere}
In this subsection, the proposed correlation is applied to model trajectories of dust particles in a high-speed flow over a sphere. Understanding high-speed particle-laden flows is relevant for entry to Mars atmosphere, where dust storms are frequent \citep{Mars_dust} and is also relevant for hypersonic flight within Earth's atmosphere, where particulates, droplets, or ice-crystals may be present. Recently, \cite{ching2020sensitivity} simulated high-speed nitrogen flow seeded with dust particles over a sphere, showing that surface heat flux is amplified by the presence of the dust particles due to interphase momentum transfer and inelastic particle-wall collisions. Interestingly, the extent of the amplification of surface heat flux was found to be sensitive to the employed drag correlations. In addition to the drag correlation, the accuracy of the surface heat flux depends on several factors, such as the employed heat flux correlation \citep{Fox_heating}, and other physical mechanisms, the details of which can be found in Ref.~\citep{ching2020sensitivity}. In this work, we perform sensitivity analysis focusing on the drag correlation, and therefore apply the proposed drag correlation to the same flow configuration considered in Ref.~\citep{ching2020sensitivity} and compare the surface heat flux obtained with other drag models. Third-order backwards difference and the third-order Adams-Bashforth method are employed to integrate the gas and particles in time, respectively.  A third-order discontinuous Galerkin (DG) schemes have been shown to have robust heat transfer predictive capability \citep{ching2019shock}. To simulate dust particles suspended in the carrier gas, a Lagrangian particle method under the DG framework \citep{ching2019submitted} has been employed. At each time step, particles are injected at random locations along the inflow boundary. Further simulation details can be found in Ref.~\citep{ching2020sensitivity}.

Figure~\ref{temp_eric} shows the temperature contour and the distribution of dust particles for a high speed nitrogen gas flow over a sphere. In Figure~\ref{heat_flux_eric}, the surface heat flux distribution obtained with the proposed drag correlation, is compared to that obtained using the Henderson and Loth models for trajectory estimation. Simulations with pure gas (i.e. no dust particles) show lower surface heat flux relative to the dusty gas. Using the proposed correlation, the heat flux predictions deviate by nearly $ \pm 14 \%$ from the Henderson ($+$) and Loth ($-$) models.

\section{Conclusions}
We develop a general drag coefficient model for spherical particles moving in a fluid, applicable for arbitrary particle relative velocity, particle diameter, gas pressure, gas temperature, and surface temperature. 
In addition to free-stream Mach number ($M_{\infty}$) and Reynolds number ($Re_{\infty}$), the proposed drag model encapsulates explicit dependence on the particle's surface temperature,  the ratio of specific heat capacities ($\gamma$), a non continuum parameter ($W_r \propto M_{\infty}^{2 \omega} / Re_{\infty}$), and the power-law index ($\omega$) of viscosity ($ \mu \propto T^{\omega}$). The correlation reproduces known theoretical results in limiting situations.
The correlation is formulated by incorporating simple physics-based scaling laws to model low-speed hydrodynamics, high-speed shock-wave physics and non continuum effects due to rarefied gas dynamics.   Free-parameters introduced in the scaling laws are obtained by using  experimental data for drag coefficients from the literature. The proposed correlation is demonstrated to be in better agreement with the experimental data compared to widely used drag models.

The dependence of the drag model on $\gamma$, and therefore on the gas-composition is investigated. The distinct trend of higher drag coefficients for monatomic gas compared to diatomic gas in experimental measurements is captured quantitatively by the proposed drag model.

Finally, the proposed drag model is applied to evaluate trajectories of particles in a simulation of high-speed dusty flow over a sphere. Enhancement of the surface heat transfer coefficient due to particle-flow and particle-surface interactions is found to be significantly different when using the new correlation compared to state-of-the-art empirical drag models.

\section*{Declaration of Interests}
Declaration of Interests. The authors report no conflict of interest.
\section*{Acknowledgments}
This work is supported by ONR FY2020 MURI grant N00014-20-1-2682, NASA NSTRF grant 80NSSC19K1129, and NASA Early Career Faculty grant (NNX15AU58G) from the NASA Space Technology Research Grants Program.

\appendix
\section{Mathematical details related to the transformations used in derivation of the drag model}
In this section, we provide additional algebraic details required in the derivation of the proposed drag model. 

\subsection{Mapping weakly compressible boundary layer to incompressible boundary layer variables \label{append:SWTtoIncomp}}
 \cite{howarth1948concerning} and \cite{stewartson1949correlated} introduced transformation to map a compressible laminar boundary layer to an equivalent incompressible boundary layer. The required transformed variables, denoted by $\tilde{...}$, in terms of free-stream variables can be expressed as:
\begin{equation}
\begin{split}
    \frac{\tilde{T}}{T_\infty} = 1+(\gamma-1)\frac{M_\infty^2 }{2} ; 
  \hspace{0.1in}   \frac{\tilde{\rho}}{\rho_\infty} = \left(\frac{\tilde{T}}{T_{\infty}}\right)^{1/(\gamma-1)}  \hspace{1.0in} 
      \label{free_stream_comp}
\end{split}
\end{equation}
Using the transformed variables along with mass conservation ($\rho_{\infty} U_{\infty} = \tilde{\rho}_{\infty} \tilde{U}_{\infty}$), $Re_{\infty}$ can be transformed to $\tilde{Re}$:
\begin{equation}
    \tilde{Re}(M_\infty,Re_\infty)= Re_\infty \left(\frac{\tilde{T}}{T}\right)^{\frac{2-\gamma}{\gamma-1}} \frac{\tilde{T}+S}{\tilde{T}+S}
    \label{Re_comp}
\end{equation}
where $S$ is the parameter in the Sutherland law of viscosity. 


\subsection{Transformation of variables across a shock-wave in the supersonic regime  \label{append:supersonic_regime}}
Post-shock conditions across a normal shock can be obtained via Rankine-Hugoniot \cite{rankine1870xv,hugoniot1887memoir} conditions which are:
\begin{equation}
    \frac{T_{s}}{T_{\infty}} = \frac{[(\gamma-1)M_\infty^2+2][2\gamma M_\infty^2-(\gamma-1)]}{(\gamma+1)^2M_\infty^2}
    \label{T_post_shock}
\end{equation}
\begin{equation}
    \frac{p_s}{p_\infty} = \frac{2 \gamma}{\gamma+1} M_{\infty}^2 -\frac{\gamma-1}{\gamma+1}
    \label{pressure_post_shock}
\end{equation}
\begin{equation}
\begin{split}
     \frac{u_s}{u_\infty} = \cfrac{2+(\gamma-1)M_\infty^2}{(\gamma+1) M_\infty^2}
     \label{u_post_shock}
\end{split}
\end{equation}
Using the jump conditions, post-shock $M_s$ can be written as:
\begin{equation}
    M_s= \sqrt{\frac{(\gamma-1)M_\infty^2+2}{2\gamma M_\infty^2-(\gamma-1)}}
    \label{Ma_post_shock}
\end{equation}
To obtain Reynolds number ($\tilde{Re_s}$), the post-shock Reynolds number ($Re_s$) is first obtained using the Rankine-Hugoniot conditions, and then transformed using the proposed transformation for curvature effects to obtain $Re_s^{\text{ef}}$. 
\begin{equation}
   Re_s^{\text{ef}} = Re_\infty \left(\frac{T_{\infty}}{\alpha^2 T_s}\right)^\omega
    \label{Re_post_shock}
\end{equation}
The second step is to employ the  Stewartson-Howarth transformation for weak compressibility effects corresponding to the post-shock Mach number ($M_s$) using Eq.~\ref{tilde_Re}: 
\begin{equation}
\begin{split}
    Re_s= Re_s^{\text{ef}}\ \Theta(M)^{\cfrac{\gamma+1}{2\gamma}-\cfrac{\gamma-1}{\gamma}\omega}
    \\
    =  Re_\infty \left(\frac{T_{\infty}}{\alpha^2 T_s}\right)^\omega
    \label{Re_post_shock}
\end{split}
\end{equation}

\subsection{Drag coefficient in the hypersonic limit} {\label{append:hypersonic}}
%
 The additional variables ($U_s/U_{\infty}, M_s^2, \Theta$, and $\alpha$) in the limit of $M_{\infty} \gg 1$ required in Eq.~\ref{cd_cont_1} to estimate $C_1$  are
\begin{equation}
    \cfrac{U_s}{U_{\infty}}\Big|_{M_\infty \gg  1} = \cfrac{\gamma -1}{\gamma +1}
    \label{UsUinfty_Mainfty}
\end{equation}
\begin{equation}
    M_s^2|_{M_\infty \gg 1} = \cfrac{\gamma -1}{2\gamma}
     \label{Mas_Mainfty}
\end{equation}
\begin{equation}
    \Theta|_{M_\infty \gg 1} = \Big[1+(\gamma-1)\cfrac{M_s^2}{2}\Big]^{\gamma/(\gamma-1)} = \Big[1+\cfrac{(\gamma-1)^2}{4\gamma}\Big]^{\gamma/(\gamma-1)}
     \label{Theta_Mainfty}
\end{equation}
\begin{equation}
    \alpha|_{M_\infty \gg 1} = \cfrac{1}{\alpha_0M_\infty}
     \label{alpha_Mainfty}
\end{equation}
Employing variables from Eq.~\ref{UsUinfty_Mainfty} to \ref{alpha_Mainfty} in the expression for $C_d$ in Eq.~\ref{cd_cont_1} yields:
 \begin{equation}
\begin{split}
     C_d^{\text{c}} (M_s,Re_s)|_{M_\infty \gg 1} \rightarrow C_{1}\Big(1-\alpha_{\infty}\cfrac{\gamma-1}{\gamma+1}\Big) \\
     +A_{o}\Big[1+\cfrac{(\gamma-1)^2}{4\gamma}\Big]^{\gamma/(\gamma-1)} = C_d^{M_\infty} \\ \implies 
     C_{1} = \cfrac{C_d^{M_\infty}-C_{o}\Big[1+\cfrac{(\gamma-1)^2}{4\gamma}\Big]^{\gamma/(\gamma-1)}}{1-\cfrac{1}{\alpha_0\ M_\infty}\cfrac{\gamma-1}{\gamma+1}}
\end{split}
\end{equation}

\subsection{Slip effects and their relevance to drag  \label{append:rarefaction}}
In view of our approach (shown in Fig.~\ref{rarefied_sphere}), we present an intuitive argument supporting Eq.~\ref{rarefied_correction} in the slip flow regime. The effect of velocity slip at the wall can be understood by considering a sphere of smaller radius which has no slip at the wall. The drag coefficient ($ C_{d_s} $) for a sphere with slip boundary condition can be obtained using the drag on the sphere of smaller radius with no slip:
\begin{equation}
    C_{d_s} = C_d \left(1-\cfrac{\delta_{Kn}}{R} \right)^2
    \label{delta_Kn1}
\end{equation}
where the $\delta_{Kn}$ is the Knudsen layer thickness. A simple scaling for $\delta_{Kn}$ based on a stationary wall in terms of slip velocity ($u_s$) and velocity gradient at the wall ($(\partial u/\partial y)|_w$) can be expressed as $\delta_{Kn} = (u_s)/(\partial u/\partial y)$.
Here, $u_s$ is the slip velocity and the wall is assumed stationary. The approximation to $u_s$ using Maxwell's slip model in terms of slip coefficient ($C_{Kn}$) and the mean free-path ($\lambda$) is given by:
\begin{equation}
    u_s = C_{Kn} \lambda \cfrac{\partial u}{\partial y} \Big|_w
    \label{slip_BC1}
\end{equation}
Eqs.~\ref{delta_Kn1} and \ref{slip_BC1} can be combined to yield an approximate expression for $\delta_{Kn}$ as:
\begin{equation}
  \cfrac{\delta_{Kn}}{R} = C_{Kn} Kn
  \label{delta_Kn_R}
\end{equation}
Substituting the above expression for $\delta_{Kn}$ from Eq.~\ref{delta_Kn_R} in Eq.~\ref{delta_Kn1}, $C_d$ reduces to:
\begin{equation}
    C_d = C_d \left(1-C_{Kn} Kn \right)^2
    \label{delta_Kn}
\end{equation}
This Knudsen correction in the limit of small Knudsen number reduces to:
\begin{equation}
    C_d \approx C_d \left(1-2 C_{Kn} Kn \right)
    \label{delta_Kn_Cd}
\end{equation}
The expression for $C_d$ in Eq.~\ref{delta_Kn_Cd}  is identical to the result obtained theoretically by Basset \cite{basset1887motion}.

\section{Expressions for the Henderson and Loth models} \label{Henderson_Loth_Equation}
\textbf{Henderson model}:
\begin{equation}
\begin{split}
 C_{d}(M_{\infty},Re_{\infty},T_{w}/T_{\infty})\ = \ & 24\left[Re_{\infty}+s\left\{4.33+\left( \cfrac{3.65-1.53T_{w}/T_{\infty}}{1+0.353T_{w}/T_{\infty}}\right)\text{exp}\left( 
-0.247\cfrac{Re_{\infty}}{s}\right) \right\}\right]^{-1}
\hspace{3.0in} \\
& 
+\text{exp}\left(-\cfrac{0.5M_{\infty}}{\sqrt{Re_{\infty}}} \right)\left[ \cfrac{4.5+0.38 (0.03Re_{\infty}+0.48\sqrt{Re_{\infty}})}{1+0.03Re_{\infty}+0.48\sqrt{Re_{\infty}}}+0.1M_{\infty}^2 \right.\\ & \left. 
 +0.2M_{\infty}^8\right] +\left[ 1-\text{exp}\left(-\cfrac{M_{\infty}}{Re_{\infty}}\right)\right]0.6s 
 \\ & \hspace{3.15in}  M_{\infty} < 1 \hspace{2.0in}
\\= & 
\cfrac{0.9+\cfrac{0.34}{M^2_{\infty}} 
+1.86\left(\cfrac{M_{\infty}}{Re_{\infty}}\right)^{1/2}
\Big[2+\cfrac{2}{s} +\cfrac{1.058}{s}
\left(\cfrac{T_{w}}{T_{\infty}}\right)^{1/2}-\cfrac{1}{s^4}\Big]}
{1+1.86\left(\cfrac{M_{\infty}}{Re_{\infty}}\right)^{1/2}}
\\
&  \hspace{3.15in} M_{\infty} > 1.75 
\\
= & C_{d}(1,Re_{\infty})+\cfrac{4}{3}(M_{\infty}-1)\left[ C_{d}(1.75,Re_{\infty})-C_{d}(1,Re_{\infty})\right]  \\
& \hspace{3.00in} 1 \geq M_{\infty} \geq 1.75 
\end{split}
\end{equation}

\textbf{Loth model}:
\begin{equation}
    C_{d} = \cfrac{24}{Re_{\infty}}\left[1+0.15Re_{\infty}^{0.687}\right]H_{M}+\cfrac{0.42C_{M}}{1+\cfrac{42500G_{M}}{Re_{\infty}^{1.16}}} \hspace{0.25in} Re_{\infty} > 45 
\end{equation}
\begin{equation}
\begin{split}
    C_{M} = \cfrac{5}{3}+\cfrac{2}{3}\tanh\left[{3\ln{\left(M_{\infty}+1\right)}}\right] \hspace{2in} M_{\infty} \leq 1.45 \\
    C_{M} = 2.044+0.2\exp\left[{-1.8\ln{\left( M_{\infty}/1.5 \right)}^2}\right]      \hspace{1.4in} M_{\infty} \geq 1.45
\end{split}
\end{equation}
\begin{equation}
\begin{split}
    G_{M} = 1-1.525M_{\infty}^4  \hspace{2.75in} M_{\infty} < 0.89
    \\
    G_{M} = 0.0002+0.0008\tanh\left[{12.77(M_{\infty}-2.02)}\right] \hspace{1.1in} M_{\infty} \geq 0.89
\end{split}
\end{equation}
\begin{equation}
   H_{M} = 1-\cfrac{0.258C_{M}}{1+514G_{M}}
\end{equation}
\begin{equation}
    C_{d} = \cfrac{C_{d,Kn,Re}}{1+M_{\infty}^4} +\cfrac{M_{\infty}^4C_{d,fm,Re}}{1+M_{\infty}^4} \hspace{0.25in} Re_{\infty} \leq 45 
\end{equation}
\begin{equation}
    C_{d,Kn,Re} = \cfrac{24}{Re_{\infty}}\left( 1+0.15Re_{\infty}^{0.687}\right)f_{Kn}
\end{equation}
\begin{equation}
    f_{Kn} = \cfrac{1}{1+Kn_{\infty}[2.514+0.8\exp{(-0.55/Kn_{\infty})}]} 
\end{equation}
\begin{equation}
    C_{d,fm,Re} = \cfrac{C_{d,fm}}{1+\left(\cfrac{C'_{d,fm}}{1.63}-1 \right)\sqrt{\cfrac{Re_{\infty}}{45}}}
\end{equation}
\begin{equation}
    C_{d,fm}' = \cfrac{(1+2s^2) \text{exp}(-s^2)}{s^3\sqrt{\pi}}+\cfrac{(4s^4+4s^2-1)\text{erf}(s)}{2s^4}
\end{equation}
\begin{equation}
   C_{d,fm} =\cfrac{(1+2s^2) \text{exp}(-s^2)}{s^3\sqrt{\pi}}+\cfrac{(4s^4+4s^2-1)\text{erf}(s)}{2s^4}+\cfrac{2}{3s}\sqrt{\pi \cfrac{T_p}{T_\infty}}
\end{equation}

\section{ Complete equations for the proposed drag model} \label{app:proposedmodel}
In this section, we summarize the full set of mathematical expressions required to calculate the drag coefficient.
\begin{equation}
    C_d = C_d^{\text{c}} (M_{\infty}, Re_{\infty}) f_{Kn_{\infty},W_r}\cfrac{1}{1+Br^\eta} + Cd_{fm}\cfrac{Br^\eta}{1+Br^\eta}
\end{equation}
\begin{equation}
\begin{split}
   Br = W_r^T\cfrac{M_{\infty}^{2\omega-1}+1}{M_{\infty}^{2\omega-1}}
    \end{split}
\end{equation}
\begin{equation}
    W_r^T = W_r \left(1+\cfrac{T_p}{T_s}\right)^\omega
    \hspace{0.5in} M_{\infty} > 1
\end{equation}
where
\begin{equation}
   W_r = \cfrac{M_{\infty}^{2\omega}}{Re_{\infty}} = Kn_{\infty} \ M_{\infty}^{2\omega-1} \sqrt{\cfrac{2}{\pi \gamma}}
 \end{equation}
  \begin{equation}
    f_{Kn_{\infty}, Wr} = \cfrac{1}{1+Kn_{\infty}\left[A_1+A_2\exp(-A_3/Kn_{\infty})\right]} \cfrac{1}{1+\alpha_{hoc} W_r^T}
\end{equation}

\begin{equation}
   C_{d,fm} =\cfrac{(1+2s^2) \text{exp}(-s^2)}{s^3\sqrt{\pi}}+\cfrac{(4s^4+4s^2-1)\text{erf}(s)}{2s^4}+\cfrac{2}{3s}\sqrt{\pi \cfrac{T_p}{T_\infty}}
\end{equation}

\begin{equation}
   s = M_{\infty} \sqrt{\frac{\gamma}{2}}
\end{equation}

\begin{equation}
\begin{split}
      C_d^{\text{c}}  = \left[C_1 \left(1-\alpha \cfrac{U_s}{U_{\infty}}\right) \right]+ \left[C_0 \Theta (M_s)\right] \left(1+\cfrac{\delta_0}{(\tilde{Re_s})^{1/2} } \right)^2 
\end{split}
\end{equation} 
\begin{equation}
\Theta (M) =  \left[ 1+(\gamma-1)\cfrac{M^2 }{2} \right]^{\gamma/(\gamma-1)}
\end{equation}
\begin{equation}
\tilde{Re}_s = Re_{\infty}\left[\cfrac{1}{\alpha^2}\cfrac{T_\infty}{ T_s}\right]^\omega \Theta(M)^{\cfrac{\gamma+1}{2\gamma}-\cfrac{\gamma-1}{\gamma}\omega}
\end{equation}
 \begin{equation}
\begin{split}
     C_{1} = \cfrac{C_d^{M_\infty}-A_{o}\Big[1+\cfrac{(\gamma-1)^2}{4\gamma}\Big]^{\gamma/(\gamma-1)}}{1-\cfrac{1}{\alpha_0 M_\infty}\cfrac{\gamma-1}{\gamma+1}}
\end{split}
\end{equation}
\begin{equation}
   \alpha = \cfrac{1}{\alpha_0M_\infty+1-\alpha_0}
\end{equation}

\bibliographystyle{jfm}
\bibliography{jfm-instructions}

\end{document}